\documentclass[review]{elsarticle}
\usepackage{graphicx}
\usepackage{amssymb,amsmath,array}
\usepackage{verbatim}
\usepackage{booktabs}
\usepackage{wasysym}
\usepackage{hyperref}
\usepackage{subcaption}
\usepackage{todonotes}
\usepackage{tabularx}
\usepackage{hhline}
\usepackage{colortbl}
\usepackage{sidecap}
\usepackage{dsfont}
\usepackage{siunitx}
\usepackage{placeins}
\usepackage{makecell}
\usepackage{makecell}
\usepackage{afterpage}
\usepackage[super]{nth}
\definecolor{Gray}{gray}{0.9}
\usepackage{floatpag}
\usepackage{datetime}

\usepackage{hyperref}

\usepackage{layouts}
\usepackage{soul}

\journal{Journal of Neurocomputing}

\begin{document}

\begin{frontmatter}

\title{Galaxy classification: \\ 
  A machine learning analysis of GAMA catalogue data}


\renewcommand*{\today}{19 January 2019}

\author[Bernoulli]{Aleke~Nolte\corref{mycorrespondingauthor}}
\cortext[mycorrespondingauthor]{Corresponding author}
\ead{a.f.nolte@rug.nl}

\author[Kapteyn,SRON]{Lingyu~Wang}

\author[Leiden,Warsaw]{Maciej~Bilicki}

\author[Leiden,USA]{Benne~Holwerda}

\author[Bernoulli]{Michael~Biehl}

\address[Bernoulli]{Bernoulli Institute for Mathematics, Computer Science and Artificial Intelligence, University of Groningen, P.O. Box 407, 9700 AK Groningen, The~Netherlands}
\address[Kapteyn]{Kapteyn Astronomical Institute, University of Groningen, Landleven 12,  9747~AD~Groningen, The~Netherlands }
\address[SRON]{SRON Netherlands Institute for Space Research, The~Netherlands}
\address[Leiden]{Leiden  Observatory,  Leiden  University,  P.O.  Box  9513, 2300~RA~Leiden, The~Netherlands}
\address[Warsaw]{Center for Theoretical Physics, Polish Academy of Sciences, al. Lotnik\'{o}w 32/46, 02-668, Warsaw, Poland}
\address[USA]{Department of Physics and Astronomy, 102 Natural Science Building, University~of~Louisville, Louisville KY~40292, USA}

\begin{abstract}
We present a 
machine learning analysis of five labelled galaxy catalogues from the Galaxy And Mass Assembly (GAMA): The \emph{SersicCatVIKING} and \emph{SersicCatUKIDSS} catalogues containing morphological features, 
the \emph{GaussFitSimple}
catalogue containing spectroscopic features,  the \emph{MagPhys} catalogue including
physical parameters for galaxies, and the \emph{Lambdar} catalogue, which contains photometric measurements.
Extending work previously presented at the ESANN 2018 conference -- in an analysis based on Generalized Relevance Matrix Learning Vector Quantization and Random Forests -- we find that neither the data from the
individual catalogues nor a combined dataset based on all 5 catalogues fully supports the  visual-inspection-based galaxy classification scheme employed to categorise the galaxies. 
In particular, only one class, the \emph{Little Blue Spheroids},
is consistently separable from the other classes. 
To aid further insight into the nature of the employed visual-based classification scheme  with respect to physical and morphological features, we present the galaxy parameters that are discriminative for the achieved class distinctions.
\end{abstract}


\end{frontmatter}


\section{Introduction}

Telescope images of galaxies reveal a multitude of appearances, 
ranging from smooth elliptical galaxies, through disk-like galaxies with spiral arms, to more irregular shapes.
The study of morphological galaxy classification plays an important role in astronomy:
the frequency and spatial distribution of galaxy types provide valuable information 
for the understanding of galaxy formation and evolution \cite{buta2013planets,mo2010galaxy}.\\
The assignment of morphological classes to  observed galaxies 
is a task which is commonly handled by astronomers.
As manual labelling of galaxies is time consuming and expert-devised classification schemes may be subject to cognitive biases, machine learning techniques have great potential to advance
astronomy by: 1)~investigating automatic classification strategies,
and 2)~by evaluating to which extent existing classification schemes
are supported by the observational data.\\
In this work, we extend a previous analysis \cite{nolteprototype} to 
make a contribution along both lines by analysing several galaxy catalogues which have been annotated using a  recent classification scheme proposed by Kelvin et al. \cite{kelvin2014galaxy}. 
In our previous study, we assessed whether this scheme is consistent with a  galaxy catalogue containing 42 astronomical parameters from the \emph{Galaxy And Mass Assembly} (GAMA, \cite{gama2009}) by performing both an unsupervised and a supervised analysis with prototype-based methods.
We assessed whether class structure can be recovered by a clustering of the data generated by the unsupervised Self-Organizing Map (SOM) \cite{kohonen1998self}, and  investigated if the morphological classification can be reproduced by   Generalized Relevance Matrix Learning Vector Quantization (GMLVQ) \cite{schneider2009adaptive}, a powerful supervised prototype-based method \cite{biehl2016prototype} \textcolor{black}{chosen for its capability to not only provide classification boundaries and class-representative prototypes, but also feature relevances.}
\textcolor{black}{
Finding consistently negative results for the supervised and unsupervised method, namely an intermediate classification accuracy of GMLVQ of around 73\%  and no clear-cut agreements between galaxy classes and SOM-clustering results, we concluded the classification scheme to  be not fully supported by the considered galaxy catalogue.}
As discussed previously \cite{nolteprototype} the hypothesised misalignment between galaxy data and classification scheme could  be explained  by  lack of discriminative power of the employed classifiers or clustering methods, by mis-labellings of certain galaxies (a possibility already discussed in \cite{Lingyu2017}), or by the absence of essential parameters in the data set.
\textcolor{black}{In this work, we address two of the mentioned aspects:  We employ an additional established and flexible classifier, Random Forests \cite{breiman2001random} to collect evidence that the previously found moderate classification performance is not due to shortcomings of GMLVQ.
Furthermore, we address} the potential incompleteness of the previously analysed dataset by performing another set of supervised analyses on several additional galaxy catalogues from the GAMA survey \cite{liske2015galaxy}, which contain a  multitude of additional  photometric, spectroscopic and morphological measurements.\\
Despite the commonly quoted abundance of data in astronomy, well-accepted benchmark datasets are not readily available in the field of galaxy classification, and only a few works analysing GAMA catalogues with machine learning methods exist.
In an analysis by Sreejith et al. \cite{Lingyu2017}, 10 features from GAMA catalogues are hand-selected and analysed using Support Vector Machines, Decision Trees, Random Forests and a shallow  Neural Network architecture. With respect to Kelvin~et.~al's classification scheme a maximum classification accuracy of 76.2\% is reported. Turner~et~al. \cite{turner2018reproducible} perform an unsupervised analysis of five hand-selected features from GAMA catalogues using k-means clustering.
While not the main aim of Turner et al.'s analysis, a comparison of the determined clusters with class information from Kelvin et al. shows galaxies that are assigned the same class by Kelvin et al. spread over several clusters (Figures 11, 13, 15 and 17 in \cite{turner2018reproducible}).
\\
In agreement with our previous results and the analyses from the above mentioned literature, we find the employed classification scheme to not be fully supported \textcolor{black}{ even when considering the additional catalogues and an alternative classifier}.
Interestingly, analogous to our previous work \cite{nolteprototype}, the \emph{Little Blue Spheroids}, a galaxy class newly introduced 
in \cite{kelvin2014galaxy}, 
remains most clearly pronounced, 
also for the set of catalogues analysed in this work.
We present the parameters that are the most relevant for the achieved class distinctions.\\

The paper is organised in as follows:
In Section~\ref{sec:data} the analysed galaxy catalogues and their preprocessing is described.
Section~\ref{sec:methods_gmlvq} outlines the employed classification methods, \textcolor{black}{GMLVQ and Random Forests}.  
Section~\ref{sec:experiments_gmlvq} describes experimental setups and results.
The work closes with a discussion in Section~\ref{sec:discussion}.\\
This paper constitutes an extension of our contribution to the \nth{26} European Symposium on Artificial Neural Networks, Computational Intelligence and Machine Learning (ESANN) 2018  \cite{nolteprototype}.
Parts of the text have been taken over literally without explicit notice. This concerns, among others, parts of the introduction and the description of GMLVQ in Section~\ref{sec:methods_gmlvq}.

\section{Data}
\label{sec:data}
In this work we analyse data from  five  galaxy catalogues (Table~\ref{tab:catalogues}) containing 
features which have been derived from spectroscopic and photometric observations, i.e. measurements of flux intensities in different wavelength bands from the  Galaxy And Mass Assembly (GAMA) survey \cite{liske2015galaxy} for a sample of 1295 galaxies.
As the catalogues contain information for different sets of galaxies, our data set consists of the set of galaxies for which a full set of features is available after balancing the relevant classes (cf. Section~\ref{sec:samp_slct}).\\
To determine this set, 
each catalogue is first cross-referenced with the galaxy sample analysed  in our ESANN contribution \cite{Lingyu2017,nolteprototype}, which contains class labels for 7941 astronomical objects. The resulting subsample is further preprocessed by selecting measurements based on the specifics of each catalogue.
Subsequently,  missing values are treated by first removing feature dimensions with a considerable amount of missing values (more than 500 missing values per feature dimension) and then discarding samples which contain missing values in any of the remaining feature dimensions.\\ 
Details of each catalogue as well as specific processing steps are delineated in the following paragraphs.

\begin{table}
\small
\centering
\begin{tabularx}{\textwidth}{llX}
\toprule
catalogue & shorthand & number of samples after preprocessing \\
\midrule
GaussFitSimple & GFS   & 7430 galaxies with 59 emission line features\\
Lambdar & Lambdar &    7365 galaxies with  28 flux measurements and uncertainties for different bands\\
MagPhys & MagPhys &    7541 galaxies with 171 features \\
SersicCatVIKING & Viking &  5476 galaxies with  66 S\'ersic features \\
SersicCatUKIDSS & Ukidss &   3008 samples with  53 S\'ersic features\\
\midrule
\makecell[l]{Complete information \\ from all catalogues} &  &  2117 galaxies \\
\midrule
\makecell[l]{Final sample \\ (cf. Section~\ref{sec:samp_slct})} & & 1295 galaxies \\
\bottomrule
\end{tabularx}
\caption{Overview of galaxy catalogues analysed in this work. Shown are also the number of samples for which complete information, i.e. information from each of the catalogues, is available, and the number of samples in the final dataset considered in the remainder of this work.} 
\label{tab:catalogues}
\end{table}

\subsection{GaussFitSimple}
The GaussFitSimple catalogue (GFS) \cite{gordon2016galaxy} contains parameters of Gaussian fits to 12 important emission lines found in galaxy spectra, namely the emission lines of oxygen 
([O I] emission lines at 6300\si{\angstrom} and
6364\si{\angstrom}, in the following denoted as \emph{OIB} and \emph{OIR},   [O II]  lines at 3726\si{\angstrom} and 3729\si{\angstrom}, denoted as \emph{OIIB} and \emph{OIIR},  [O III] lines at 4959\si{\angstrom} and
5007\si{\angstrom}, denoted as \emph{OIIIR} and \emph{OIIIB}), nitrogen ([N II] lines at 6548\si{\angstrom} and
6583\si{\angstrom}, \emph{NIIR} and \emph{NIIB}), sulphur ([S II] lines at 6716\si{\angstrom} and 6731\si{\angstrom}, \emph{SIIR} and \emph{SIIB}), and hydrogen (H$\alpha$ and H$\beta$  lines at
6563\si{\angstrom}  and  
4861\si{\angstrom}, respectively).
Further, the catalogue contains slope and intercept of the continuum, that is, the background radiation in-between emission lines.
In addition to these parameters the catalogue also contains meta-information concerning model fits and corresponding errors.\\
From the GaussFitSimple catalogue we select amplitudes (AMP\_*) and sigma (SIG\_*) of the Gaussian fit for each emission line, as well as calculated fluxes (*\_FLUX) and equivalent widths (*\_EW). Here and in the following, the asterisk * is a placeholder for  the name of the corresponding emission line. 
We further include information about the continuum (CONT, GRAD) and the strength of the D4000 break, resulting in 59 selected features.
We discard all samples for which a failure of the fitting procedure has been indicated (FITFAIL\_*), 
and remove samples containing missing values in any of the feature dimensions. 
The resulting sub-catalogue then contains 7430 galaxies with 59 emission line features. \\
We note that the classification performance on the full catalogue, which contains model fit information and 
errors / measurement uncertainties is comparable to the results achieved with the reduced catalogue containing 59 features (cf. Section~\ref{sec:experiments_gmlvq}).
As the selected parameters allow for a more direct interpretation in terms of emission line strengths
and therefore facilitate interpretation from the astronomical perspective, we consider the reduced catalogue in the following.

\subsection{Lambdar}
\label{sec:lambdar}
The Lambdar catalogue \cite{wright2016galaxy_lambdar} contains flux measurements and uncertainties for 21 bands, as measured by the LAMBDAR software
\cite{wright2016galaxy_lambdar}. 
When cross-referencing with the catalogue analysed in our preceding study, 
400 galaxies are missing from the Lambdar catalogue.
These galaxies are removed from the considered Lambdar subset and do not contribute to the ensuing missing
value calculations.
Columns still containing a considerable amount of missing values after this step ($>$ 500 )
are excluded from the analysis.
The removed columns contain parameters that include fluxes and errors in the far and near Ultraviolet (UV) (FUV\_flux, FUV\_fluxerr, NUV\_flux, NUV\_fluxerr), and fluxes and errors in the 100$\mu m$ to 500$\mu m$ bands (P100\_flux, P100\_gcfluxerr, 
P160\_gcflux, P160\_gcfluxerr, S250\_gcflux, S250\_gcfluxerr, S350\_gcflux, S350\_gcfluxerr, 
 S500\_gcflux, and  S500\_gcfluxerr). 
After removing these, 28 features remain in the catalogue, namely fluxes and errors for u, g, r, i and z bands observed in the Sloan Digital Sky Survey (SDSS,\cite{york2000sloan}), Z, Y, J, H and K bands from  VISTA Kilo-Degree Infrared Galaxy Survey (VIKING, \cite{edge2013vista}), and W1, W2, W3 and W4 bands from the Wide-field Infrared Survey Explorer (WISE, \cite{wright2010wise}).
After this step, samples that are  missing measurements for any of the remaining features
are removed, resulting in a final sub-catalogue of 7365 galaxies with  28 features.

\subsection{MagPhys}
The MagPhys catalogue \cite{da2008simple_magphys} contains physical parameters comprising information about stellar populations as well as parameters describing the inter-stellar medium in the galaxies.
Parameters include, among others, star formation rates, star formation time-scales, information about star formation bursts, as well as the masses of stars formed in the bursts, overall stellar ages and masses, metallicities, and information about dust in the interstellar medium and in stellar birth clouds ; all this for each included galaxy.
All MagPhys parameters have been derived from information provided in the  Lambdar catalogue (Section~\ref{sec:lambdar}) using the  MAGPHYS program \cite{da2008simple_magphys}.
Due to missing values in the Lambdar catalogue, the MagPhys catalogue does not contain information for  400 of the galaxies analysed in our ESANN contribution \cite{nolteprototype}.
Apart from these, there are no missing values, 
so that information from 177 MagPhys features is available for 7541 galaxies.
However, after selecting the final sample (cf. Section~\ref{sec:samp_slct}) some parameters exhibit almost no variance over the considered samples: 
Parameters \emph{fb17\_percentile2\_5, fb18\_percentile2\_5, fb17\_percentile16, fb17\_percentile50, fb17\_percentile84 and fb18\_percentile16}\footnote{Percentiles of the likelihood distribution of parameters
describing  the fraction of the effective stellar mass  formed in bursts over the last $10^7$ and $10^8$ years}
are largely constant, with maximally 15 data points displaying deviations. We therefore remove these features, which results in a dimensionality of 171 for the final MagPhys sample.\\
Information on the MagPhys parameter shorthand notation used in the remainder can be found in \cite{magphys}.

\subsection{S\'ersic Catalogues}
Three different catalogues are available which contain parameters of 
single-S\'ersic-component fits to the 2D surface brightness distribution of galaxies in different bands \cite{kelvin2012galaxy}.
The single-S\'ersic-component fits have been produced with the  GALFIT program \cite{peng2002detailed_galfit}.
The catalogues contain a parameter, GALPLAN\_*,  which indicates GALFIT fitting failures for each band, where the asterisk * is a placeholder for the band.
GALPLAN\_*$=$0 indicates a severe failure when fitting the surface brightness profile of the galaxy, which could not be amended by attempting a number of correction strategies. We therefore discard all samples where GALPLAN\_*$=$0.\\
An additional goodness-of-fit parameter allowing to judge the quality of profile fitting is the PSFNUM\_* parameter. This parameter indicates the number of prototype stars used to model the point spread function (PSF) in the galaxy image to which  the surface brightness profile was fit.
As indicated in the GAMA catalogue description,  modelling PSFs based on less than 10 stars may result in poor PSF models, which in turn may result in poorly fitted surface brightness distributions.
Accordingly, we discard all samples where the PSFNUM\_* parameters have a value lower than 10.\\
The catalogue further contains meta-information needed to reproduce the results of the GALFIT fitting.
Here we concentrate on parameters that are descriptors of galaxies as opposed to parameters describing the fitting procedure.
The  galaxy descriptors, all GALFIT-derived, are:
GALMAG\_*, the  
magnitude of the S\'ersic model; 
GALRE\_*, the  
half-light radius measured along the semi-major axis; 
GALINDEX\_*, the  
S\'ersic index; 
GALELLIP\_*, the 
ellipticity;
GALMAGERR\_*, the  
error on magnitude;
GALREERR\_*, the  
error on the half-light radius; 
GALINDEXERR\_*, the  
error on the S\'ersic index; 
GALELLIPERR\_*, the  
error on ellipticity; 
GALMAG10RE\_*, the  
magnitude of a model truncated at 10 $\times$ the half-light radius; 
GALMU0\_*,  the 
central surface brightness; 
GALMUE\_*,  the 
effective surface brightness at the half-light radius; 
GALMUEAVG\_*,  the 
effective surface brightness within the half-light radius;
and
GALR90\_*,  the 
radius containing 90\% of total light, measured along the semi-major axis of the galaxy.

\subsubsection{SersicCatVIKING}
The SersicCatVIKING \cite{kelvin2012galaxy} catalogue contains the above  measurements for the VIKING bands Z, Y, J, H, and K.
Based on the GALFIT failure parameter GALPLAN\_*$=$0, 966 samples were removed from the sub-catalogue.
Additional 1074 samples were removed because of PSFNUM\_* $< 10$.
After removing samples which have missing values in any of the named feature dimensions the final  sub-catalogue
contains 5476 galaxies with  66 S\'ersic features.

\subsubsection{SersicCatUKIDSS}
The SersicCatUKIDSS \cite{kelvin2012galaxy} catalogue contains the above  measurements for the UKIDSS \cite{ukidss2007} bands Y, J, H, K.
Based on the GALFIT failure parameter GALPLAN\_*$=$0, 2904 samples were removed from the sub-catalogue.
Additional 1841 samples were removed because of PSFNUM\_* $< 10$.
After removing samples which have missing values in any of feature dimensions the final  sub-catalogue
contains 3008 samples with  53 S\'ersic features.

\subsubsection{SersicCatSDSS}
For the SersicCatSDSS catalogue \cite{kelvin2012galaxy}, most samples from the cross-referenced catalogue \cite{nolteprototype,Lingyu2017} are discarded based on the PSFNUM and GALPLAN selection, and only 1672 samples remain.
The SersicCatSDSS catalogue is therefore excluded from the analysis.

\subsection{Classification Scheme}
For each galaxy analysed in our ESANN contribution \cite{nolteprototype}, a class label has been determined by astronomers following a visual inspection based classification scheme described by Kelvin et al. \cite{kelvin2014galaxy}.
The scheme assigns galaxies to 9 classes:
\emph{Ellipticals}, 
\emph{Little Blue Spheroids},
\emph{Early-type spirals},
\emph{Early-type barred spirals},
\emph{Intermediate-type spirals},
\emph{Intermediate-type  barred spirals}, 
\emph{Late-type spirals \& Irregulars}, 
\emph{Artefacts} and  
\emph{Stars} (Table~\ref{tab:9classes}).
We will refer to the classes by their class index (1-9).\\
As barred spirals, artefacts and stars are highly under-represented in this sample, our subsequent analysis will
focus on the substantial classes, namely classes 1, 2, 3, 5 and 7.
\begin{table}
\small
\centering
\begin{tabular}{lllr}
\toprule
\makecell[l]{class index} & class name & \makecell[l]{corresponding\\ Hubble type} &  \makecell[l]{prevalence in \\ data set of \cite{nolteprototype, Lingyu2017}}\\
\midrule
\rowcolor{Gray}
1 & Ellipticals &  E0-E6 & 11\% \\
\rowcolor{Gray}
2 & Little blue Spheroids & -   & 11\% \\
\rowcolor{Gray}
3 & Early-type spirals  & S0,  Sa & 10\% \\
4 & Early-type barred spirals  & SB0,  SBa & 1\%\\
\rowcolor{Gray}
5 & Intermediate-type spirals & Sab, Scd & 15\%\\
6 & Intermediate-type  barred spirals  & SBab, SBcd & 2\%\\
\rowcolor{Gray}
7 & Late-type spirals \& Irregulars & Sd - Irr & 45\%\\
8 & Artefacts & - & 0.4\% \\
9 & Stars     & - & 0.005\%\\
\bottomrule
\end{tabular}
\caption{Overview of galaxy classes in the dataset used to cross-reference the catalogues analysed in this paper.
Shown are also the corresponding Hubble types, an established galaxy type descriptor in astronomy, and the class index that is used to identify classes in the remainder of the work.
Gray highlights indicate the classes that are part of the final classification problems.}
\label{tab:9classes}
\end{table}

\subsection{Sample selection}
\label{sec:samp_slct}
To ensure a fair comparison between the catalogues, our final dataset comprises the subsample of galaxies for which a full set of measurements is available, i.e galaxies for which measurements are provided in each of the five considered  catalogues.
This is the case for 2117 galaxies.
Considering only the substantial classes 1, 2, 3, 5 and 7, and balancing classes so that for each class the same number of samples is selected, (259, based on class 2, the class with minimum cardinality), results in a final sample of 1295 galaxies.

\section{Methods: Classifiers}
\label{sec:methods_gmlvq}
\subsection{GMLVQ}
Generalized Relevance Matrix LVQ (GMLVQ) \cite{schneider2009adaptive,biehl2016prototype} is an extension of Learning Vector Quantization (LVQ) \cite{kohonen1997learning}.\
LVQ is a supervised prototype-based method, in which prototypes are annotated with a class label.
The prototypes are 
adapted based on the label information of the training data: if the best-matching unit (BMU), the prototype closest to the data point,  is of the same class as a given data point, the prototype is moved
towards the data point, while in the case of a BMU with an incorrect class label, the prototype is repelled. 
While LVQ assesses similarities between prototypes and data points using the Euclidean distance, GMLVQ learns a distance measure that is tailored to the data, allowing it to suppress noisy feature dimensions or to emphasise distinctive features and their pair-wise combinations.
GMLVQ therefore considers a generalized distance 
\begin{equation*}
\textstyle
d^\Lambda (\mathbf{w}, \boldsymbol{\xi}) = (\boldsymbol{\xi} - \mathbf{w})^T \Lambda \,\,(\boldsymbol{\xi} - \mathbf{w}) \; 
\text{~~with } \Lambda = \Omega^T \Omega \; \text{~~and~ }
\sum_i \Lambda_{ii} = 1,
\end{equation*}
where $\Lambda$ is an $n\times n$ positive semi-definite matrix, 
$\boldsymbol{\xi} \in \mathds{R}^n$ 
represents a feature vector and $\mathbf{w} \in \mathds{R}^n $ is one of $M$ prototypes.
After optimisation, the diagonal of $\Lambda$ will encode the learned relevance of the feature dimensions, while the off-diagonal elements encode the relevances of pair-wise feature combinations. 
As empirically observed and theoretically studied \cite{biehl2015stationarity,biehl2012large} the relevance matrix after training is typically low rank and can be used, for instance, for visualisation of the data set (see ~\ref{sec:appendix_viz}
for an example).\\
The parameters $\{\mathbf{w}_i\}_{i=1}^M$ and $\Lambda$ are optimised based on a heuristic cost function, see 
\cite{schneider2009adaptive}, 
\begin{equation}
\textstyle
E_{\text{GMLVQ}} = \sum_{i=1}^P \mu_i^\Lambda, \text{ with }
   \mu_i^\Lambda = (d_J^\Lambda(\boldsymbol{\xi}_i) - 
                         d_K^\Lambda(\boldsymbol{\xi}_i)) /
                        (d_J^\Lambda(\boldsymbol{\xi}_i) + 
                         d_K^\Lambda(\boldsymbol{\xi}_i)) \,,
\end{equation}
where $P$ refers to the number of training samples,  $d_J^\Lambda(\boldsymbol{\xi}) = d_J^\Lambda(\mathbf{w}_J,\boldsymbol{\xi}) $ denotes the distance to the closest correctly labelled prototype $\mathbf{w}_J$, 
and $d_K^\Lambda(\boldsymbol{\xi}) = d_K^\Lambda(\mathbf{w}_K,\boldsymbol{\xi}) $
denotes the distance to the closest incorrect prototype $\mathbf{w}_K$.
If the closest prototype has an incorrect label, 
$d_K^\Lambda(\boldsymbol{\xi}_i)$ 
will be smaller than $d_J^\Lambda(\boldsymbol{\xi}_i)$, 
hence, the corresponding $\mu_i^\Lambda$ is positive. 
Minimisation of $E_\text{GMLVQ}$ will therefore favour
the correctness of nearest prototype classification.
In a stochastic gradient descent procedure based on a single
example the update reads
\begin{equation}
\textstyle
\mathbf{w}_{J,K} \leftarrow  \mathbf{w}_{J,K} - \eta_w \partial \mu_i / \partial \mathbf{w}_{J,K} 
\text{  and  }
\Omega \leftarrow \Omega - \eta_\Omega \partial \mu_i / \partial \Omega \;.
\label{eq:GMLVQ}
\end{equation}
Derivations and full update rules can be found in  \cite{schneider2009adaptive}.
In a batch gradient descent version \cite{gmlvqcode}, updates of the form (\ref{eq:GMLVQ}) are summed over all training samples. \\

\subsection{Random Forests}
Random Forests (RF) \cite{breiman2001random} is a well-known classification and regression method that employs an ensemble of randomised Decision Trees \cite{breiman2017classification}. 
In randomised Decision Trees,
a subset of features is chosen randomly at each node. Considering only the selected features, decision thresholds are determined based on the best attainable split between classes.
To combine the classifications of each tree in the ensemble, i.e. to determine the output of the Random Forest, different methods can be employed. In the scikit-learn implementation used in our experiments \cite{scikit-learn,scikit-RF} the final classification output is obtained by averaging the probabilistic prediction of each tree. \\
Details on the set-up of the experiments for RF as well as for GMLVQ can be found in Section~\ref{sec:setup}.

\section{Experiments}
\label{sec:experiments_gmlvq}
In our experiments, we assess relevances of features and discriminability between classes by training and evaluating GMLVQ for each of the five  preprocessed catalogues described in Section~\ref{sec:data}.
As found in previous work \cite{nolteprototype},  class 2, the Little Blue Spheroids (LBS), were particularly well-distinguishable. We perform experiments for both, the full 5-class problem, trying to distinguish between galaxy classes 1, 2, 3, 5 and 7 (cf. Table~\ref{tab:9classes}) and a 2-class problem in which the LBS are classified against galaxies from the other four classes.
In addition to the single catalogue experiments, we also assess feature relevances and discriminability between classes for a concatenation of all catalogues, to account for possible synergies between features from different  catalogues.\\
To allow for interpretation in the light of other classifiers, we perform the same experiments with the widely used Random Forests (RF) classifier \cite{breiman2001random} as a baseline.

\subsection{Setup}
\label{sec:setup}
We train and evaluate GMLVQ on the galaxy catalogue data using a publicly available implementation \cite{gmlvqcode}.
As the GMLVQ cost function is implicitly biased towards classes with larger numbers of samples, we  train and evaluate the classifier on size-balanced random subsets of the five classes.
For our experiments, we specify one prototype per class and run the algorithm 
for 100 batch gradient steps with step size adaptation as realised in \cite{gmlvqcode} with default parameter settings.the 
We validate the algorithm by performing a class-balanced \emph{repeated random sub-sampling validation} (see e.g. \cite{friedman2001elements} for validation methods) for a total of 10 runs.
Error measures and relevance profiles shown in the following correspond to averages over the 10 repetitions.
For the two-class problems we also obtain and average Receiver Operator Characteristics (ROC) and the 
corresponding Area under the Curve (AUC) \cite{fawcett2006introduction}.

\subsubsection{Setup LBS vs others}
For the two-class problem, we evaluate the classifier on a subset of the full dataset (cf. Section~\ref{sec:samp_slct}) containing 515 samples. 
For this subset, we select all 259 samples from class 2, while the \emph{others} class is made up by 256 samples consisting of  64 samples  randomly selected from class 1, 3, 5, and 7 each.
The remaining settings and validation procedure remain identical to the 5-class problem.

\subsubsection{Random Forests}
\label{sec:rf_setup}
We execute experiments employing Random Forests analogous to the  GMLVQ experiments, i.e. the classifier is trained on class-balanced random subsets of the data and validated using repeated random sub-sampling validation.
Experiments are performed using a publicly available scikit-learn implementation \cite{scikit-learn,scikit-RF} with default settings. 

\subsection{Classification results based on parameters from individual catalogues}
\begin{figure}
\centering
    \begin{subfigure}[b]{1\textwidth}
    \includegraphics[width=1\textwidth]{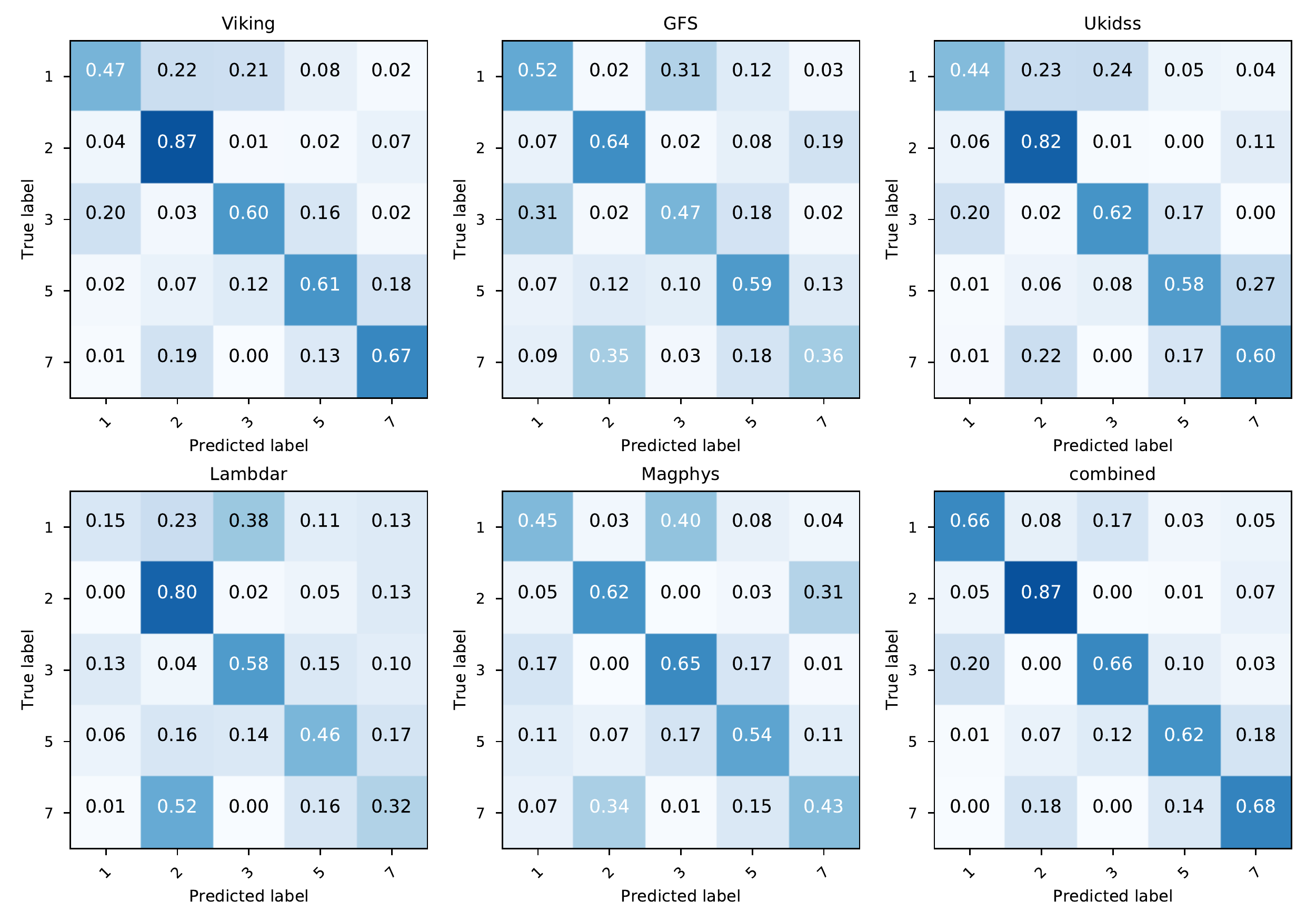}
    \caption{5 class problem: Confusion matrices for GMLVQ performance for each of the 5 single catalogues and the combined catalogue.}
    \label{fig:confmats}
    \end{subfigure}

    \begin{subfigure}[t!]{0.45\textwidth}
	\centering
	\includegraphics[width=\textwidth]{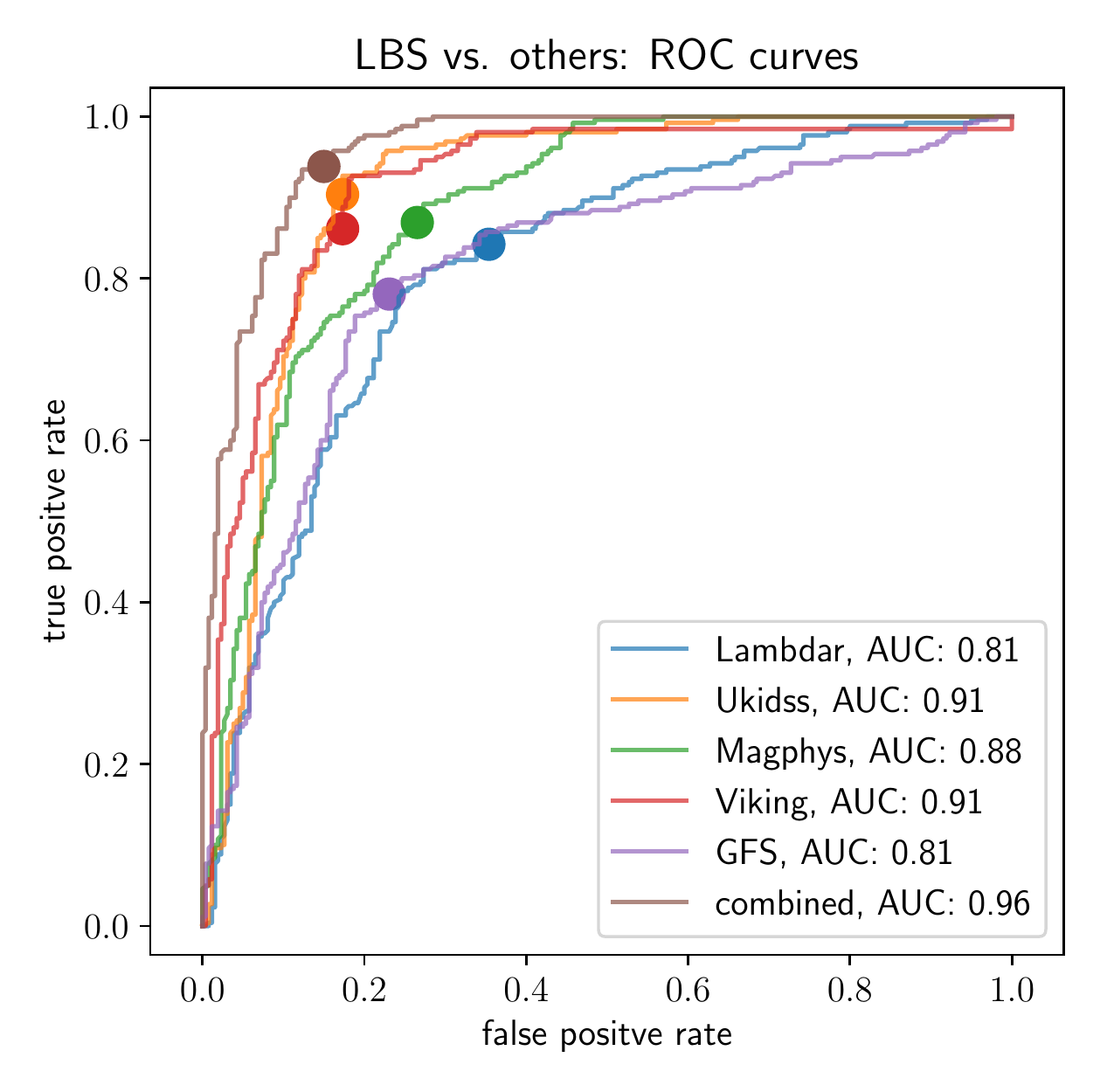}
        \caption{2 class problem: ROC curves for each of the 5 single catalogues and the combined catalogue.}
        \label{fig:roc}
    \end{subfigure} \hfill
    \begin{subfigure}[t!]{0.45\textwidth}
	\includegraphics[width=\textwidth]{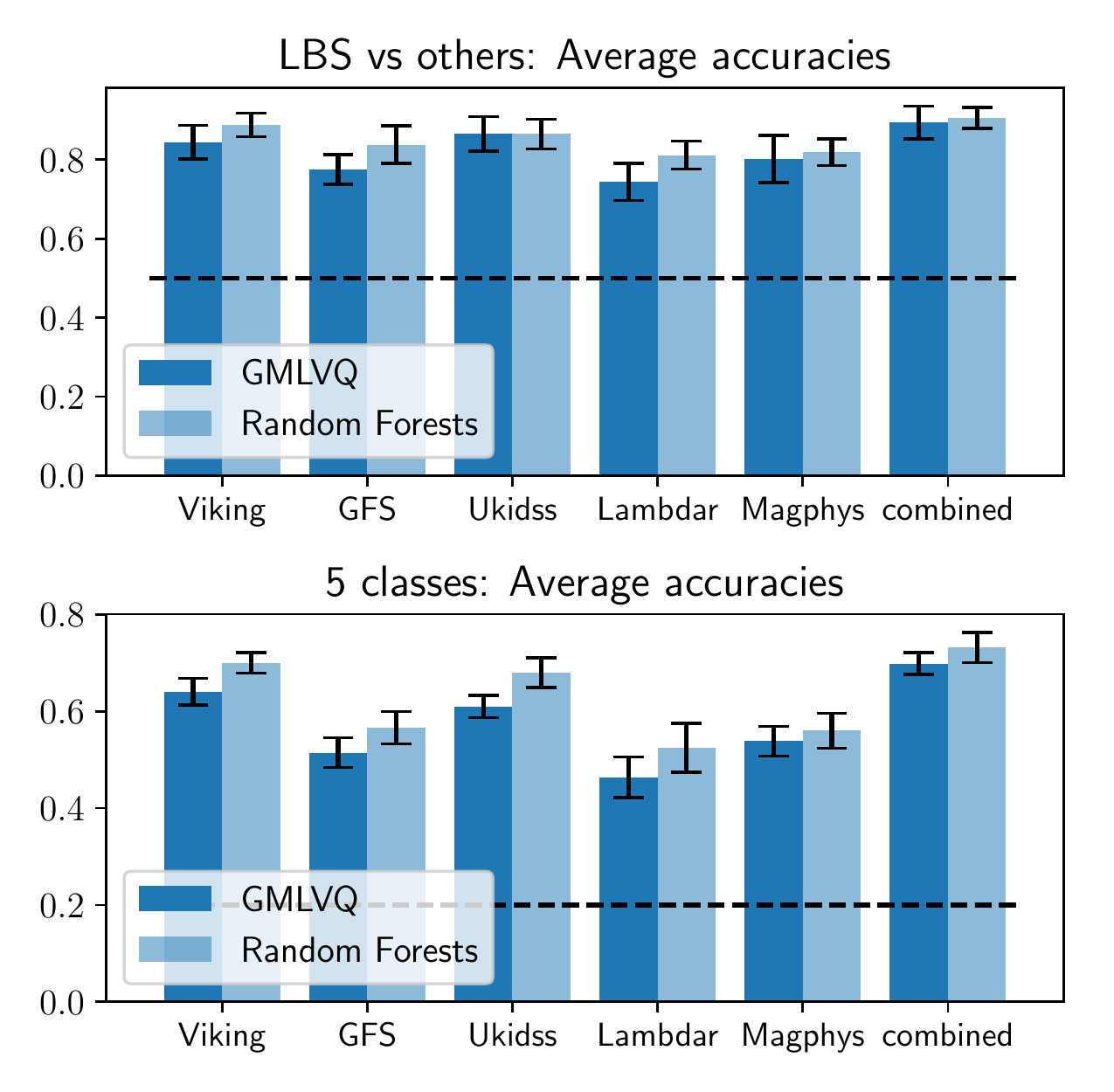}
	\caption{GMLVQ classification accuracy for the different catalogues for both the 2-class and 5-class problem. For comparison, classification accuracies achieved by Random forests are shown-side by side. Note that the combined condition for the Random Forests considers all available parameters, while for GMLVQ a relevance-based pre-selection is made. Dashed lines indicate chance level performance.}
        \label{fig:accuracies}
\end{subfigure}
\caption{Summary of GMLVQ classification performances for both single and combined catalogues.}
\label{fig:performances}
\end{figure}

A summary of classification performances for both the 5-class and the 2-class problem can be found in Figure~\ref{fig:performances}.
For the 5-class problem, an overview of confusion matrices (averaged over all validation runs) for each of the catalogues is shown in Figure~\ref{fig:confmats}; an overview of the average classification accuracies can be found in Figure~\ref{fig:accuracies} in the bottom panel.
For the 2-class problem, a comparison of ROC curves and classification accuracies can be found in Figure~\ref{fig:roc} and  in Figure~\ref{fig:accuracies} in the top right subfigure, respectively.\
The corresponding average relevance profiles  contrasting feature relevances for the  5-class and 2-class problem are shown in the Appendix, in
Figure~\ref{fig:lambdar} (Lambdar catalogue), 
Figure~\ref{fig:gfs} (GaussFitSimple catalogue), 
Figure~\ref{fig:viking} (SersicCatVIKING catalogue),
Figure~\ref{fig:ukidss} (SersicCatUKIDSS catalogue),
and
Figures~\ref{fig:magphys_5class} and~\ref{fig:magphys_LBS} (MagPhys catalogue).

\paragraph{Results based on SersicCatVIKING}
The confusion matrix indicating the GMLVQ class-wise accuracy on the SersicCatVIKING catalogue exhibits similar, albeit slightly worse performance  than the performances presented 
in our previous work \cite{nolteprototype} that was based on  a different set of galaxy parameters.
Based on the SersicCatVIKING, the  LBS are classified with higher accuracy (87\% vs. 91\% in ESANN) than the other classes (47-67\%, 64-74\%).
As in the ESANN results, classes 1 and 3 show some overlap (21\% of class 1 samples are classified as class 3, and 20\% of class 3 samples are erroneously classified as class 1). 
However, unlike in the ESANN results, the overlap between class 1 and class 2 is increased in the classification using SersicCatVIKING:  22\%  of class 1 samples are now classified as belonging to class 2, where this overlap was only 10\% for the data analysed in our ESANN contribution \cite{nolteprototype}.
This is also reflected in the 2-class problem when distinguishing the LBS from the other classes.
In \cite{nolteprototype} this can be achieved with AUC(ROC)=0.96, while for the SersicCatVIKING catalogue the classification accuracy is around 84\% and the AUC(ROC)=0.91. 
Another notable increase in overlap is the overlap between class 5 and 7, where the misclassification rate of class 5 galaxies as class 7 galaxies is increased from 8\% to 18\%.

\paragraph{Results based on GaussFitSimple Catalogue}
The confusion matrix for the classification based on the GaussFitSimple Catalogue shows the highest classification accuracy of 64\% for the LBS. Class 3 drops in accuracy to  47\% . This is in part due to an increased overlap between the classes, 31\% of class 1 samples are classified as class 3 samples and 31\%  of class 3 samples as belonging to class 1. In addition, there is increased overlap between class 1 and 5 (12\%) and class 3 and 5 (18\%), while the overlap between classes 1 and 3 with both LBS and class 7 remains low.
It is notable that based on the information in the GaussFitSimple Catalogue, class 7  is only classified slighly above chance level, with most of its samples being misclassified as class 2 (35\%) and class 5 (18\%). 
Despite this, the distinction between LBS and others is still on average 78\% correct, the AUC(ROC)=81\%.

\paragraph{Results based on SersicCatUKIDSS}
The results for the SersicCatUKIDSS show an overall similar performance to the results of the SersicCatVIKING catalogue:
In comparison to the classification performance presented in our ESANN contribution \cite{nolteprototype}, there is an  increased misclassification of class 1 samples as class 2 samples, and an increased misclassification of class 5 samples as belonging to class 7.
LBS classification accuracy is at 87\% with an  AUC(ROC)=0.91.

\paragraph{Results based on Lambdar Catalogue}
The results for the Lambdar sample show a similar picture as the GaussFitSimple sample:
Class 7 is classified with an accuracy of only slightly above chance level and is often (52\%) misclassified as class 2.
Unlike in the GFS results, the accuracy for class 1 is below chance level (15\%).
As has been the case for the other catalogues, class 1 samples are misclassified mostly as class 3 (38\%).
In contrast to the GaussFitSimple catalogue, here class 1 also shows considerable overlap with class 2 (23\% of class 1 samples are misclassified as class 2). In addition, 
a considerable amount of class 1 samples (11\% and 13\%)  are also misclassified as classes 5 and 7.
Further, class 5 and class 3 show overlap, with 15-16\% misclassifications.
Overall, classification accuracy based on the Lambdar catalogue is lowest (46\%), while the LBS can be distinguished with 74\% accuracy and an AUC(ROC)=0.81 .

\paragraph{Results based on MagPhys catalogue}
The classification results for the MagPhys sample show a similar trend as the results based on the  Lambdar sample:
Classes 1 and 3 exhibit considerable overlap (40\% of class 1 samples are classified as class 3, and 17\% of class 3 samples are classified as class 1), class 7 accuracy is low (43\%) and is frequently misclassified as class 2 (34\% of the cases).
In contrast to the Lambdar sample, there is almost no overlap between class 1 and class 2.
Average classification accuracy for the 5 classes based on the MagPhys catalogue is at (54\%), while the LBS can be distinguished with 80\% accuracy and an AUC(ROC)=0.88 .

\paragraph{LBS vs other}
The LBS can be distinguished from the other classes with an intermediate accuracy of about 74\% - 87\% and AUC(ROC) values of 81\%-91\%.

\subsection{Combined catalogues}
\begin{figure}
\centering
\begin{subfigure}[b]{0.48\textwidth}
\includegraphics[width=\textwidth]{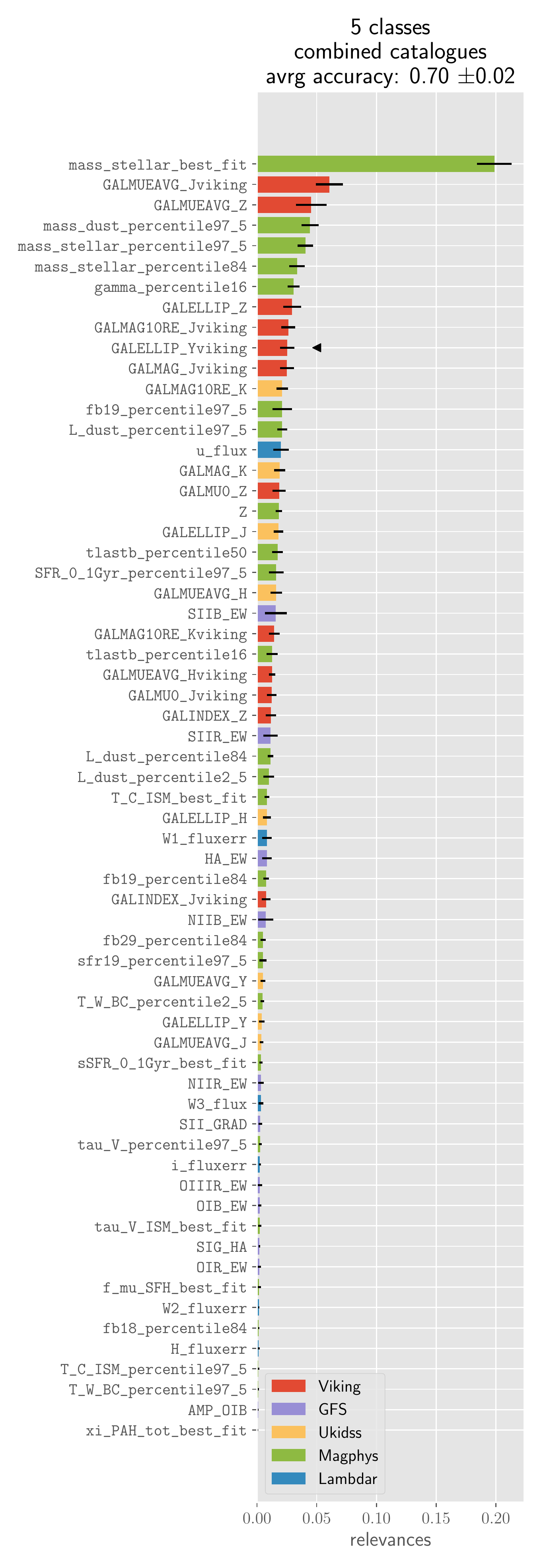}
\caption{}
\label{fig:relevances_combined_5class}
\end{subfigure}
\begin{subfigure}[b]{0.48\textwidth}
\includegraphics[width=\textwidth]{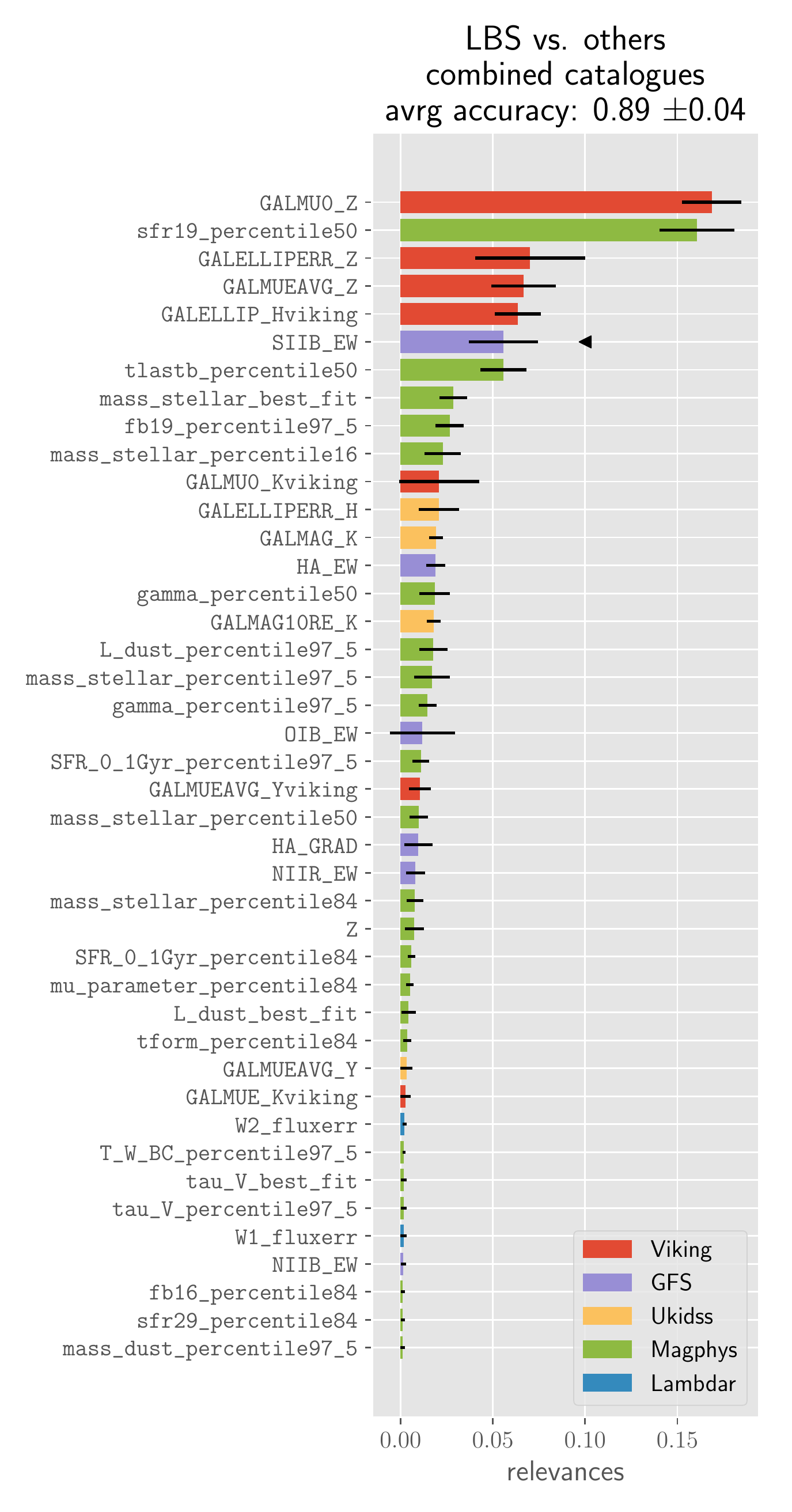}
\caption{}
\label{fig:relevances_combined_2class}
\end{subfigure}
\caption{Sorted relevance profiles for catalogues obtained by combining the  most relevant features that cumulatively make up 50\% of the relevances in the single catalogue relevance profiles. Bar colours indicate the origin catalogue for each feature.
Features up to the position marked by a black arrow constitute 50\% of the cumulative relevance determined for the resulting combined catalogue.}
\end{figure}

Combining all catalogues would result in a very high-dimensional classification problem, thereby rendering the resulting relevance profiles difficult to interpret.
We therefore  select a subset of parameters from each individual catalogue based on the feature relevances obtained in the single catalogue experiments in the following manner:
 For each individual catalogue, parameters are sorted according to their relevance. Subsequently, the most relevant parameters cumulatively comprising 50\% of the summed total relevance are carried over to the combined catalogue. We note that we have also performed GMLVQ experiments on the full catalogue comprising all 377 features, which resulted in similar, albeit slightly worse performances than reported below.\\ 
 For the Random Forests baseline experiments, we select the full catalogue of 377 features independent from the GMLVQ results, as to warrant identical experimental conditions. For completeness, we note that classification accuracy of Random Forests on the above described relevance-selected parameter subset is comparable to the classification accuracy on the full dataset.  \\
Sorted relevance-profiles for the resulting combined catalogues are displayed in Figure~\ref{fig:relevances_combined_5class} and Figure~\ref{fig:relevances_combined_2class}, for the 5-class and 2-class problem, respectively.
To simplify comparison, the confusion matrix as well as the 2-class classification performance are displayed alongside  the  individual catalogue performances in Figure~\ref{fig:performances}.\\
~
Considering the confusion matrix for the combined catalogue, 
a slight overall increase in performance with respect to the individual  catalogue performances can be observed. Further, it reflects the combined properties of the individual catalogues:
An overlap between classes 1 and 3, some overlap between class 3 and 5, and some overlap between class 2 and 7.
In comparison to the results presented in \cite{nolteprototype}, classification accuracy is slightly decreased (70\% vs. 73\%). It should be noted however, that in \cite{nolteprototype} thrice as many samples per class were available, which could account for the difference in performance.
LBS can be distinguished from the other classes with a classification accuracy of 89\% and an AUC(ROC)=0.96.

\paragraph{Feature relevances for the combined catalogues} 
The parameters that make up 50\% of the relevances for the 
5-class and the 2-class problem (indicated by a black arrow in Figure~\ref{fig:relevances_combined_5class} and Figure~\ref{fig:relevances_combined_2class}), almost exclusively originate from the SersicCatVIKING and MagPhys catalogues.
For the 5-class problems, these parameters are related to stellar masses and dust (mass\_stellar\_best\_fit, mass\_dust\_percentile97\_5, mass\_stellar\_percentile\_97\_5 and mass\_stellar\_percentile84), and  the star formation timescale (gama\_percentile16),
the effective surface brightness within
the half-light radius for the J- and Z-bands (GALMUEAVG\_J and GALMUEAVG\_Z), ellipticity of the galaxy (GALELLIP\_Z, GALELLIP\_Yviking),  and magnitude of a GALFIT model of the galaxy (GALMAG10RE\_Jviking).\\
For the 2-class problem, the most relevant parameters encompass the GALFIT central surface brightness in Z-band (GALMU0\_Z), parameters related to star formation rates (sfr19\_percentile50), information related to the ellipticity of the galaxies (GALELLIPERR\_Z, GALELLIP\_Hviking), effective surface brightness (GALMUEAVG\_Z) and information about the equivalent width of the sulphur emission line.\\
It should be noted that relevance-matrices are not necessarily unique. They depend on which other features are available and on the parameters chosen for both data preprocessing and execution of the algorithm.
This can be illustrated when considering highly correlated variables: GMLVQ might assign either two intermediate relevances to each of the variables, or deem one variable highly relevant at expense of the other correlated variable's relevance.
Relevance profiles therefore should be interpreted in the sense that focusing on the most relevant parameters would allow differentiation between classes with the reported accuracy, while keeping in mind that other combinations of features may achieve this as well.

\subsection{Random Forests baseline results}
The classification accuracies for Random Forests for the individual and combined catalogues are displayed in Figure~\ref{fig:accuracies} side-by-side with the GMLVQ results. For all catalogues applying the Random Forest classifier results in comparable, though slightly better classification accuracies.

\section{Discussion \& Conclusion}
\label{sec:discussion}
The results presented above suggest that there may be inconsistencies in 
the investigated  morphological classification scheme:
Analogous to our previous findings \cite{nolteprototype}, it has proven difficult to distinguish 
galaxy types using \textcolor{black}{two powerful and flexible classifiers,} GMLVQ and Random Forests.
In all GMLVQ analyses of the individual  as well as of the  combined catalogues, class 1 (\emph{Ellipticals}) and 3 (\emph{Early-type spirals}) are particularly difficult to differentiate.
Class 7 (\emph{Late-type spirals \& Irregulars}) is frequently misclassified as class 5 (\emph{Intermediate-type spirals})  and with a similar frequency  as class 2 (LBS), while class 2 is consistently detected with the highest sensitivity among all classes.\\
The difficulty of training a successful classifier 
was also observed in \cite{Lingyu2017}, where class-wise averaged accuracies are around 75\%.
As mentioned in our earlier contribution \cite{nolteprototype}, possible explanations for poor classification performance may be the lack of discriminative power of the employed classifiers or mis-labellings of certain galaxies \cite{Lingyu2017}. A possible indication for the latter case may be that samples from class 7 (\emph{Late-type spirals \& Irregulars}) are often misclassified as class 5 (\emph{Intermediate-type spirals}), and class 2 (LBS).
This indicates that the feature representations of the galaxies in question share more properties with the named classes, and it is not unlikely that in the hand-labelling process an \emph{Intermediate-type spiral} is occasionally misclassified as class 7 (e.g. confused with a \emph{Late-type spiral}), or that a LBS is classified as class 7 (an \emph{Irregular}).
In the former case, employing even more flexible classifiers, 
e.g. GMLVQ with local relevance matrices 
\cite{schneider2009adaptive}, may improve classification performances. 
In the second case, 
if 
mis-labellings 
are restricted to ``neighboring'' classes in an assumed underlying 
class ordering (e.g. when considering class 5 adjacent to class 7, or class 1 (\emph{Ellipticals}) as adjacent to class 3 (\emph{Early-type spirals})),
ordinal classification 
may provide further insights
\cite{fouad2012adaptive, tang2017ordinal}.\\
Despite trying to address the issue of essential parameters being not contained in the dataset analysed in  \cite{nolteprototype} by considering 5 additional catalogues with a multitude of  photometric, spectroscopic and morphological measurements,
it is still possible that additional (and possibly not yet discovered) parameters would enable improved class distinction.
Yet, our results do not rule out
the possibility that the true, underlying 
grouping of galaxies is considerably different and less clear-cut than the investigated one.
Further data-driven analyses of galaxy parameters and images with advanced clustering methods 
might reveal alternative groupings, like recently found for data in the 
VIMOS Public Extragalactic Redshift Survey \cite{siudek2018vimos},
or even suggest novel classification schemes.\\
To aid further insight into the nature of the employed visual-based classification scheme, in particular with respect to physical parameters, we have presented relevances of the catalogue features for the investigated class distinctions.
Note that relevances have to be interpreted with regard to the characteristics of the data sample (e.g. correlations) and classification performance. This connotes that feature relevances  are only meaningful when the class of interest is at least moderately well distinguished from the others. \textcolor{black}{Further it should be noted that the presented feature relevances are not necessarily unique -- alternative relevance solutions may exist.}
It is of particular interest to note that in the combined catalogue the most relevant features originate from the S\'ersic catalogues and the MagPhys catalogue. The high relevance of S\'ersic features  indicate the importance of galaxy structure
   in different bands for the class distinction, while the presence of highly relevant features  from the MagPhys catalogue highlights that classification performance is aided by these physical parameters as well.
Further insight into the role of features in the context of necessary and dispensable features may be obtained by studying feature relevance bounds along the lines of  \cite{gopfert2018interpretation}.

\paragraph{Conclusions}
We have presented an analysis of five galaxy catalogues using \textcolor{black}{Random Forests and} GMLVQ, a prototype-based classifier. Analogous to results obtained in preceding work on a lower-dimensional  dataset, we conclude that even when considering a multitude of additional galaxy descriptors,
the visual-based classification scheme used to label the galaxy sample remains not fully supported by the available data. Taking into account that perceptual and conceptual biases likely play non-negligible roles in the creation and application of galaxy classification schemes,   further data-driven analyses
might help provide  novel insights regarding the true underlying grouping of galaxies.

\paragraph{Acknowledgements}
\footnotesize{
GAMA is a joint European-Australasian project based around a spectroscopic campaign using the Anglo-Australian Telescope. The GAMA input catalogue is based on data taken from the Sloan Digital Sky Survey and the UKIRT Infrared Deep Sky Survey. Complementary imaging of the GAMA regions is being obtained by a number of in- dependent survey programmes including GALEX MIS, VST KiDS, VISTA VIKING, WISE, Herschel-ATLAS, GMRT and ASKAP providing UV to radio coverage. GAMA is funded by the STFC (UK), the ARC (Australia), the AAO, and the participating institutions. The GAMA website is \url{http://www.gama-survey.org/}.\\
We thank Sreevarsha Sreejith,  Lee Kelvin and Angus Wright for helpful feedback and discussions and the anonymous reviewers for feedback which helped us improve the manuscript.\\
A. Nolte and M. Biehl acknowledge financial support by the EU’s Horizon 2020 research and innovation programme
under Marie Sklodowska-Curie grant agreement No 721463 to the SUNDIAL ITN network. M. Bilicki is supported by the Netherlands Organization for Scientific Research, NWO, through grant number 614.001.451 and by the Polish Ministry of Science and Higher Education through grant DIR/WK/2018/12.}

\appendix
\FloatBarrier
\setcounter{figure}{0} 

\section{Dataset visualizations and intrinsic dimensionality reduction in GMLVQ}
\label{sec:appendix_viz}
\begin{figure}[h!]
\floatpagestyle{empty}
    \centering
        \includegraphics[width=1.0\textwidth]{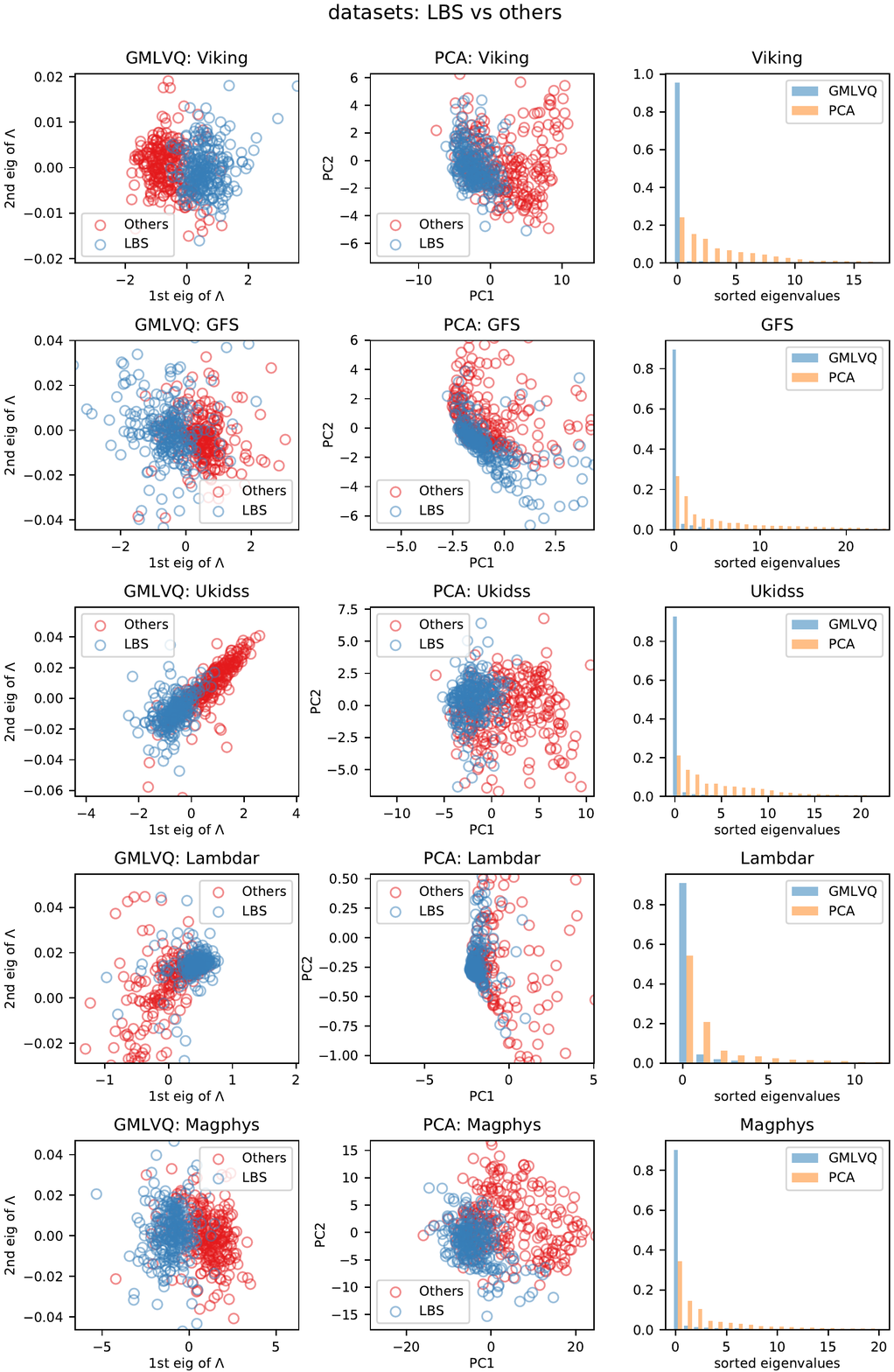}
    \caption{2D visualisations of the datasets used in the \emph{LBS vs. others} classification condition. The leftmost column displays a projection of each dataset onto the first two eigenvectors of the learned relevance matrix $\Lambda$. In the middle column, projections of the datasets onto the first two principal components (PC1 and PC2) are shown.
    The right column juxtaposes the eigenvalue spectra of the relevance matrix and the data covariance matrix used in PCA. 
    For increased readability, figures concentrate on the median region of the data and axes are  cut off at a  3 times inter-quantile range distance from the median.
    Furthermore, the data projections are scaled by the square root of the corresponding eigenvalues.
    In the sub-figures of the eigenvalue spectra the x-axis is truncated after both eigenvalues have dropped below a value of 0.005.}
        \label{fig:LBS_vs_others_data_proj}
    \end{figure}

\begin{figure}[h!]
    \centering
        \includegraphics[width=1.\textwidth]{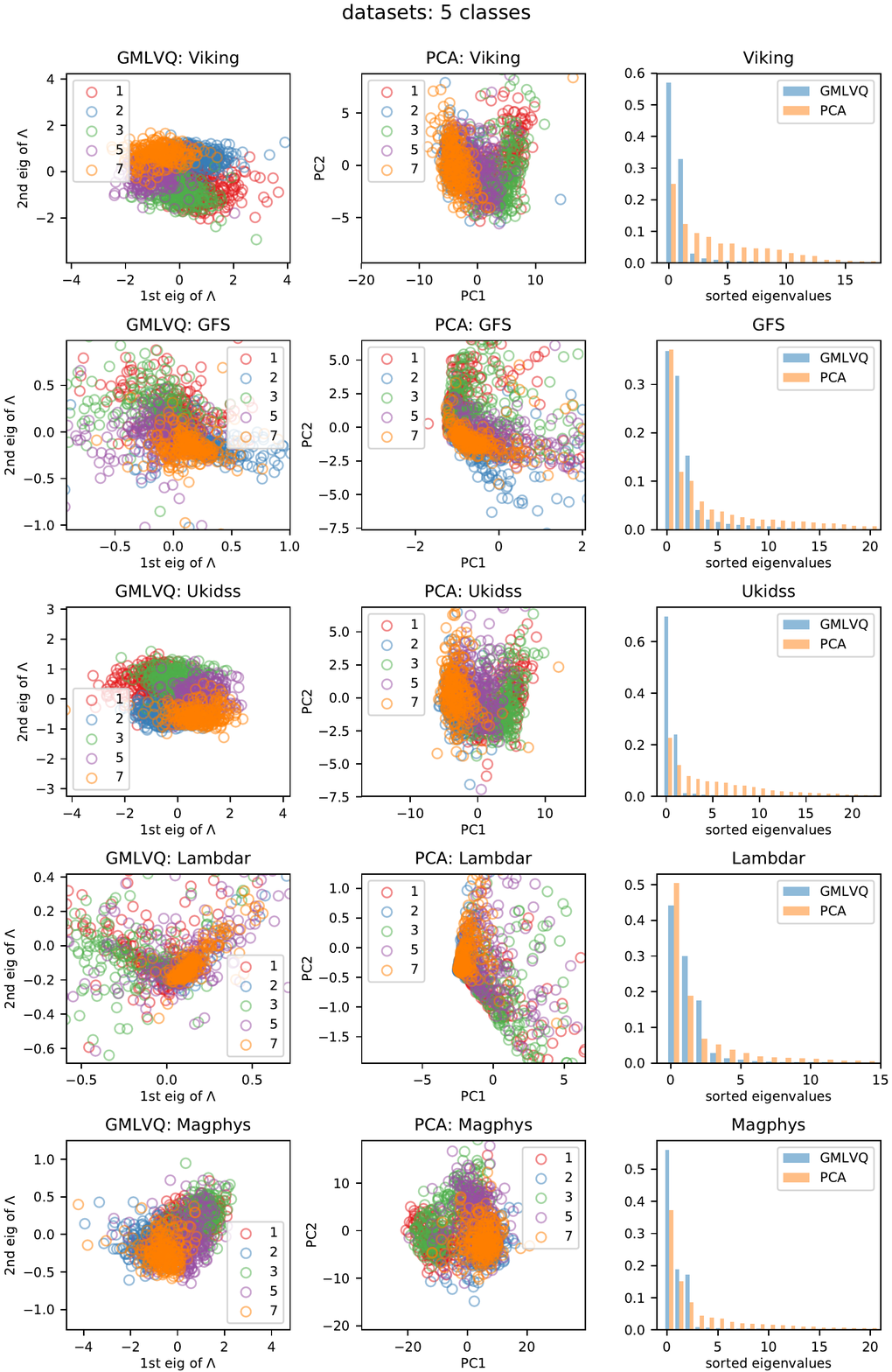}
         \caption{2D visualizations of the datasets used in the \emph{5-class} classification condition. The leftmost column displays a projection of each dataset onto the first two eigenvectors of the learned relevance matrix $\Lambda$. In the middle column, projections of the datasets onto the first two principal components (PC1 and PC2) are shown.
    The right column juxtaposes the eigenvalue spectra of the relevance matrix and the data covariance matrix used in PCA. 
    For increased readability, figures concentrate on the median region of the data and axes are  cut off at a  3 times inter-quantile range distance from the median.
    Furthermore, the data projections are scaled by the square root of the corresponding eigenvalues.
    In the sub-figures of the eigenvalue spectra the x-axis is truncated after both eigenvalues have dropped below a value of 0.005.}
        \label{fig:5class_data_proj}
    \end{figure}

Figures~\ref{fig:LBS_vs_others_data_proj} and~\ref{fig:5class_data_proj} display projections of each dataset considered in this work onto the first and second eigenvector of the relevance matrix $\Lambda$ (cf. Section~\ref{sec:methods_gmlvq})  and onto the first two principal components determined by Principal Component Analysis (PCA) \cite{jolliffe2011principal}.
The rightmost column of each figure contrasts the eigenvalue spectra  of $\Lambda$ and the data covariance matrix which forms the basis for PCA.
While $\Lambda$ is an $n \times n$ matrix, the steeply declining eigenvalue spectra for each dataset illustrate the low-dimensional subspace which GMLVQ operates in after learning \cite{biehl2015stationarity,biehl2012large}.
In particular, for the 5 class problem, $\Lambda$ spans an approximately 3 dimensional subspace, while for the 2 class problem the subspace is essentially one-dimensional.
The low-rank relevance matrices therefore can be thought of as performing a GMLVQ-intrinsic dimensionality reduction.\\
Comparing the 2-D projections  onto the two leading eigenvectors of $\Lambda$ and the projections onto the first two principal components, the former results in a more fanned out representation with respect the classes.
This is due to the fact that by making use of the class labels, GMLVQ finds a lower-dimensional discriminative subspace as opposed to the unsupervised PCA.

\FloatBarrier
\section{Feature relevances for individual catalogues}
\setcounter{figure}{0} 
\textcolor{black}{In the following (Figures~\ref{fig:lambdar}-~\ref{fig:magphys_LBS}), we present relevance profiles for the individual catalogues analysed in this work.
Relevance profiles reflect the diagonal of GMLVQ's relevance matrix $\Lambda$ after learning (cf. Section~\ref{sec:methods_gmlvq}) and summarise the importance of features for a given data sample and  classification task. Figures display mean and variance of the profiles over 10 independent runs (cf. Section~\ref{sec:setup}).
As noted previously, for an accurate interpretation it is important to note that, in general, relevance profiles are not unique: Especially in the presence of correlated variables, alternative profiles resulting in comparable classification performance might exist.
In particular, a feature's low relevance does not entail the feature to carry no information for the desired class distinction, but may instead indicate its contribution to be at least partly redundant with other features.\\
For example, contrary to expectations at first glance, our experiments with the Lambdar sample result in relevance profiles that indicate uncertainties of fluxes of various bands as more relevant than the corresponding flux measurements themselves (Figure~\ref{fig:lambdar}).
While it is not unthinkable that flux uncertainties systematically vary over a subset of galaxy classes (personal communication, Angus Wright, developer of the LAMBDAR software), in our sample W1 and W2 fluxes are correlated with both their respective errors and with fluxes from other bands. W1 and W2 fluxes as well as fluxes from other bands are thus at least partly redundant with the W1 and W2 flux uncertainties, and therefore might end up more relevant than the corresponding fluxes.
}

\begin{figure}[hb!]
    \centering
    \begin{subfigure}[b]{0.45\textwidth}
        \includegraphics[width=\textwidth]{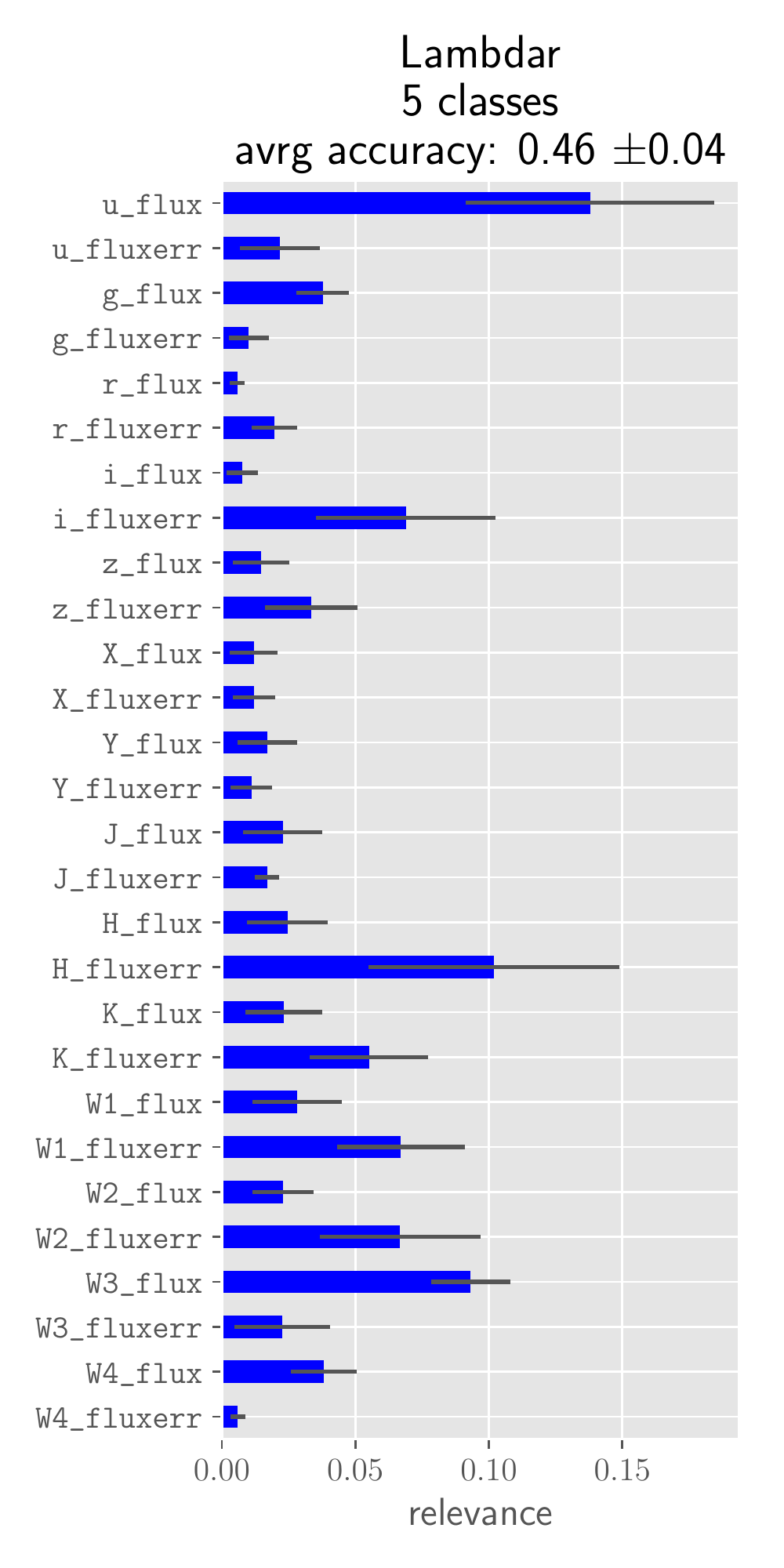}
        \label{fig:lambdar_5class}
    \end{subfigure}
    ~
    \begin{subfigure}[b]{0.45\textwidth}
        \includegraphics[width=\textwidth]{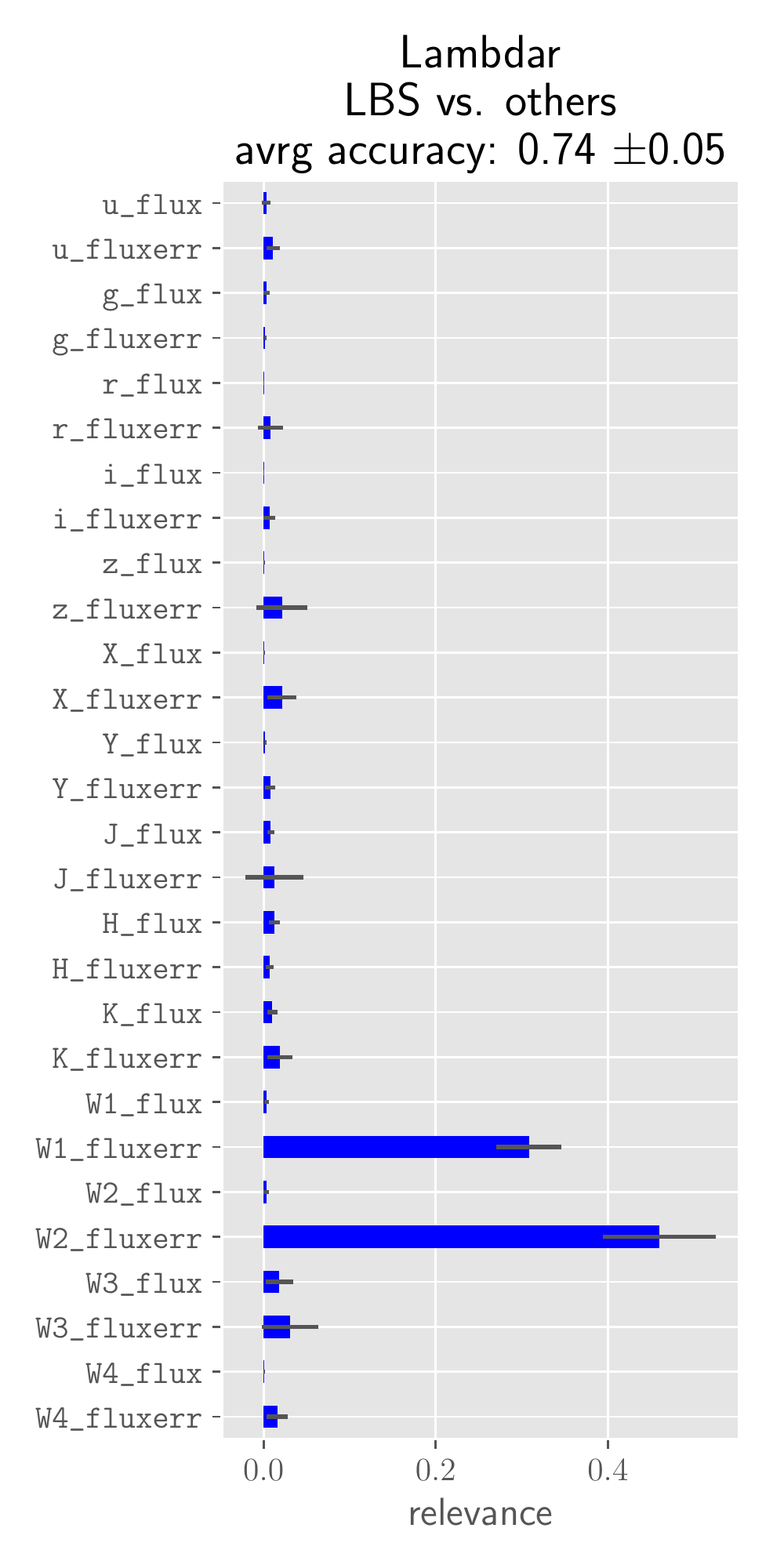}
        \label{fig:lambdar_LBS}
    \end{subfigure}
\caption{Feature relevances as determined by GMLVQ for the Lambdar sample. \textcolor{black}{
For accurate interpretation of the relevance profiles, take note that relevance profiles are not necessarily unique, in particular in the presence of highly correlated variables. This connotes that  focusing on the relevant parameters would enable to differentiate between classes with the
reported accuracy, however, there may be other combinations of features which could result in  similar accuracies.}
 }
    \label{fig:lambdar}
\end{figure}

\begin{figure}
    \centering
    \begin{subfigure}[b]{0.45\textwidth}
        \includegraphics[width=\textwidth]{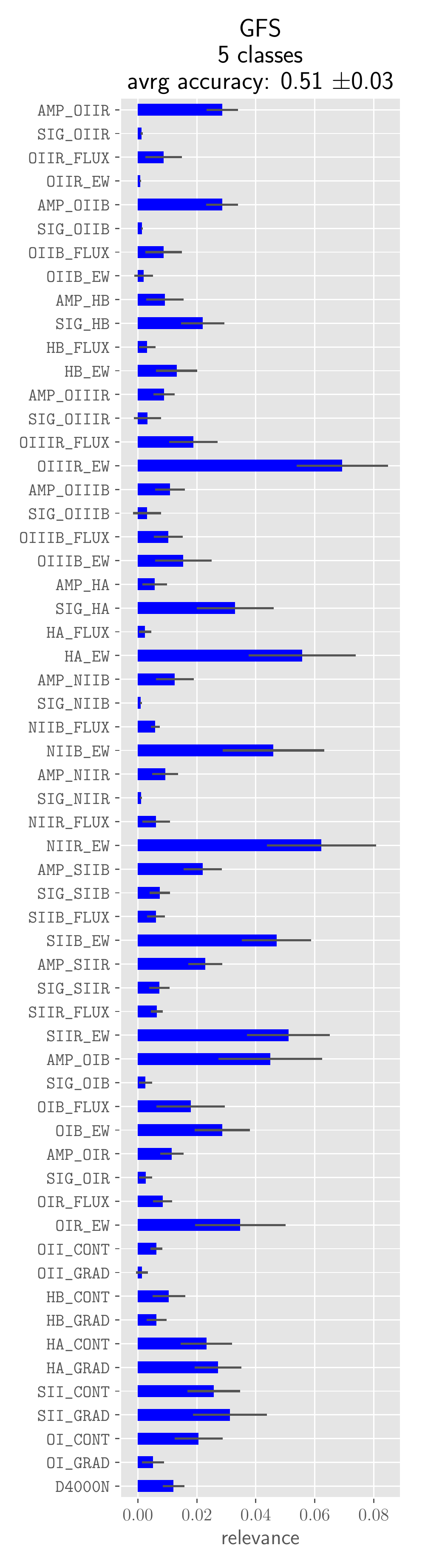}
        \label{fig:gfs_5class}
\end{subfigure}
\begin{subfigure}[b]{0.45\textwidth}
        \includegraphics[width=\textwidth]{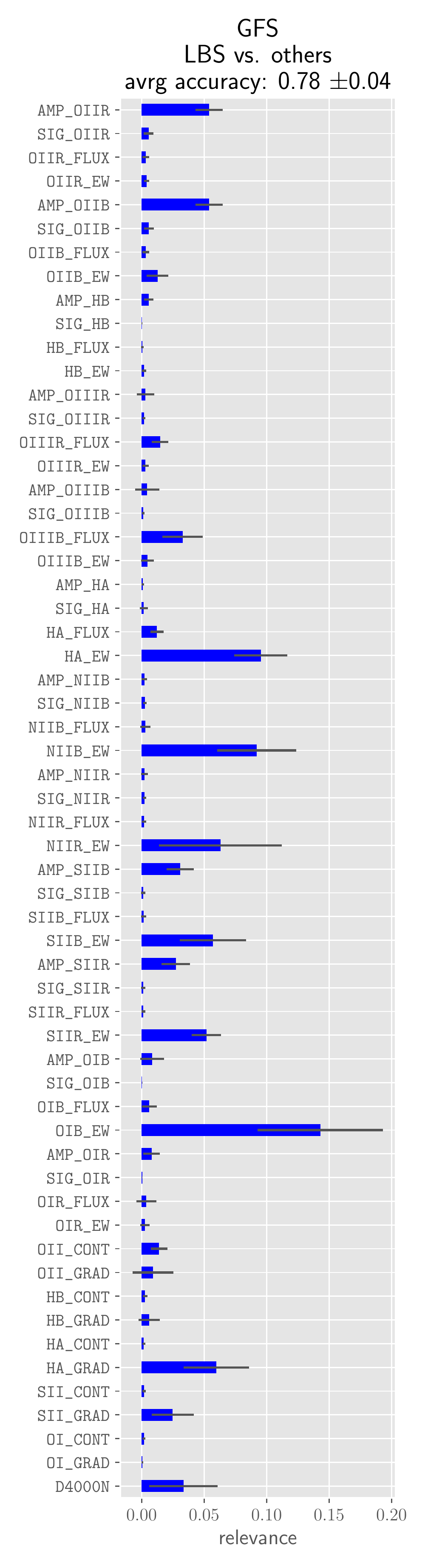}
        \label{fig:gfs_LBS}
\end{subfigure}
    \caption{Feature relevances as determined by GMLVQ for the GaussFitSimple sample. \textcolor{black}{Same note applies here as to Fig.~\ref{fig:lambdar}.}
 }
    \label{fig:gfs}
\end{figure}

\begin{figure}
    \centering
    \begin{subfigure}[b]{0.45\textwidth}
        \includegraphics[width=\textwidth]{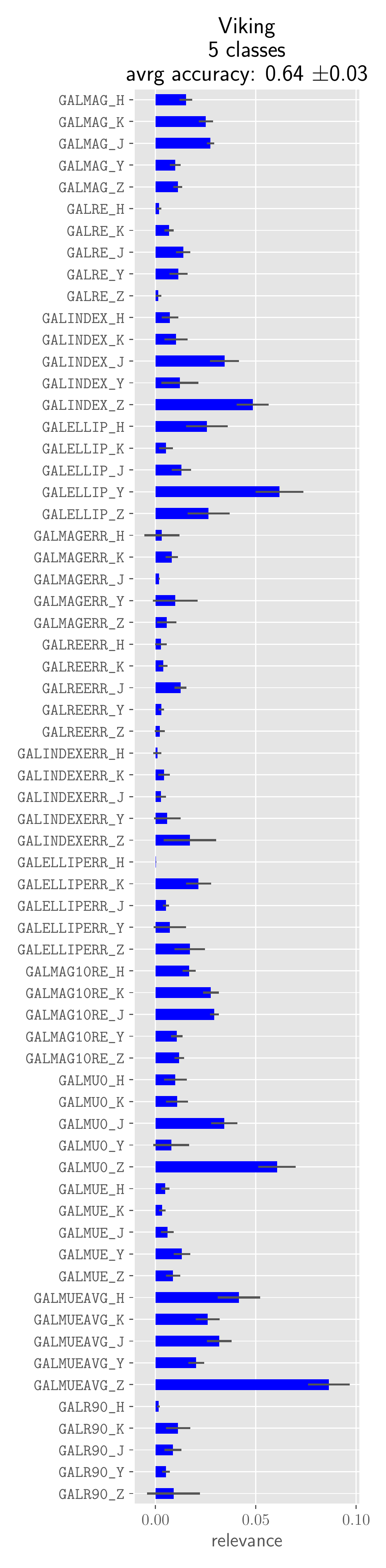}
        \label{fig:viking_5class}
    \end{subfigure}
    ~
    \begin{subfigure}[b]{0.45\textwidth}
        \includegraphics[width=\textwidth]{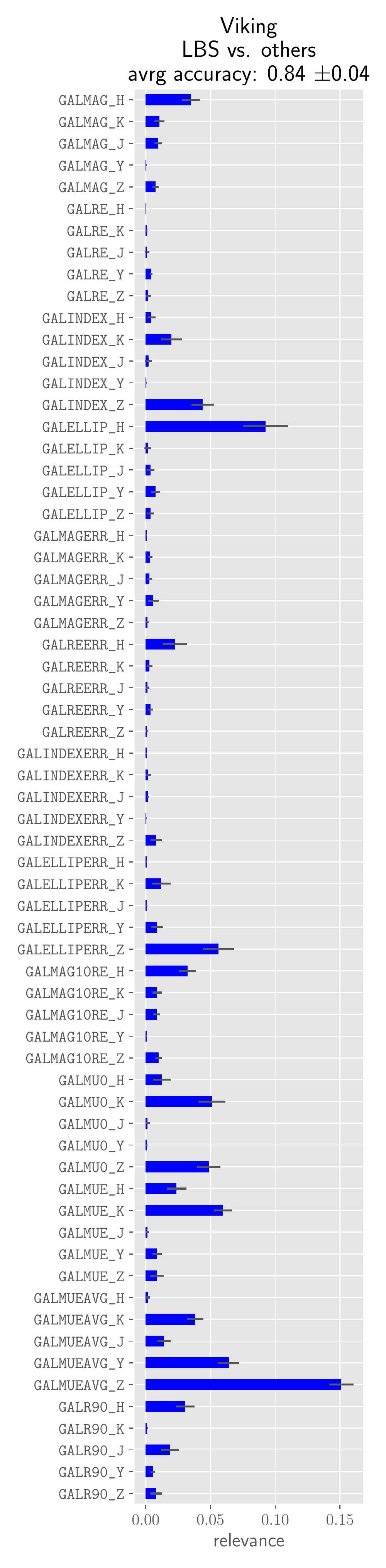}
        \label{fig:viking_LBS}
    \end{subfigure}
    
    \vspace*{-11mm}
    \caption{Feature relevances as determined by GMLVQ for the SersicCatVIKING sample.
     \textcolor{black}{Same note applies here as to Fig.~\ref{fig:lambdar}.}
 }
    \label{fig:viking}
\end{figure}

\begin{figure}
    \centering
    \begin{subfigure}[b]{0.45\textwidth}
        \includegraphics[width=\textwidth]{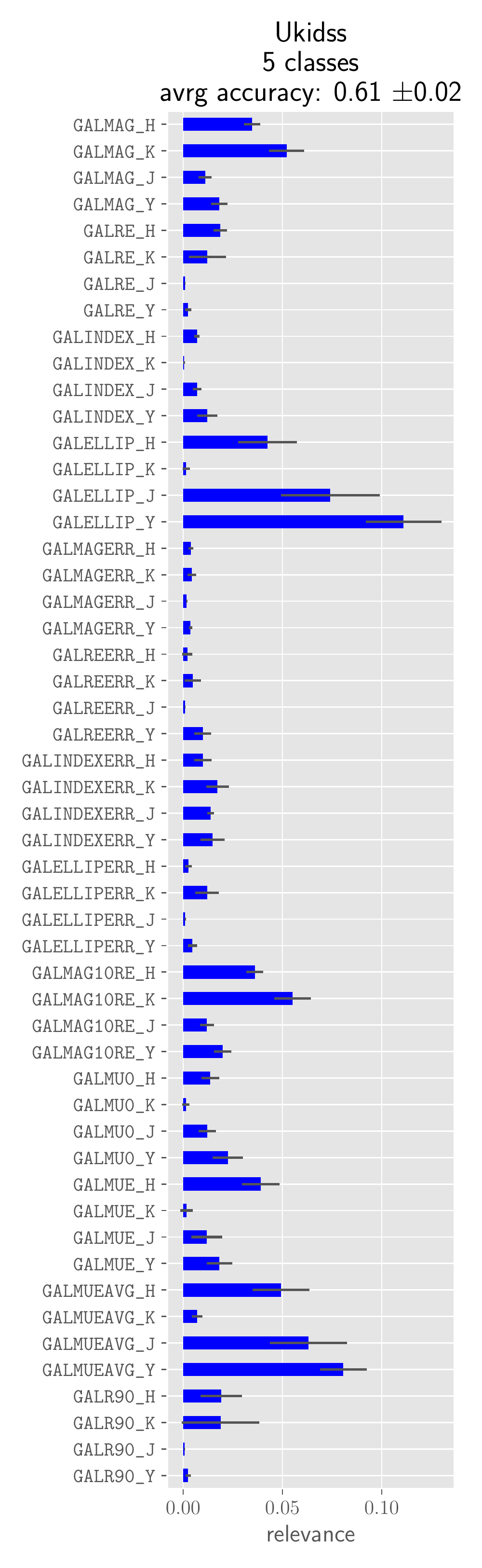}
        \label{fig:ukidss_5class}
    \end{subfigure}
    ~
    \begin{subfigure}[b]{0.45\textwidth}
        \includegraphics[width=1.1\textwidth]{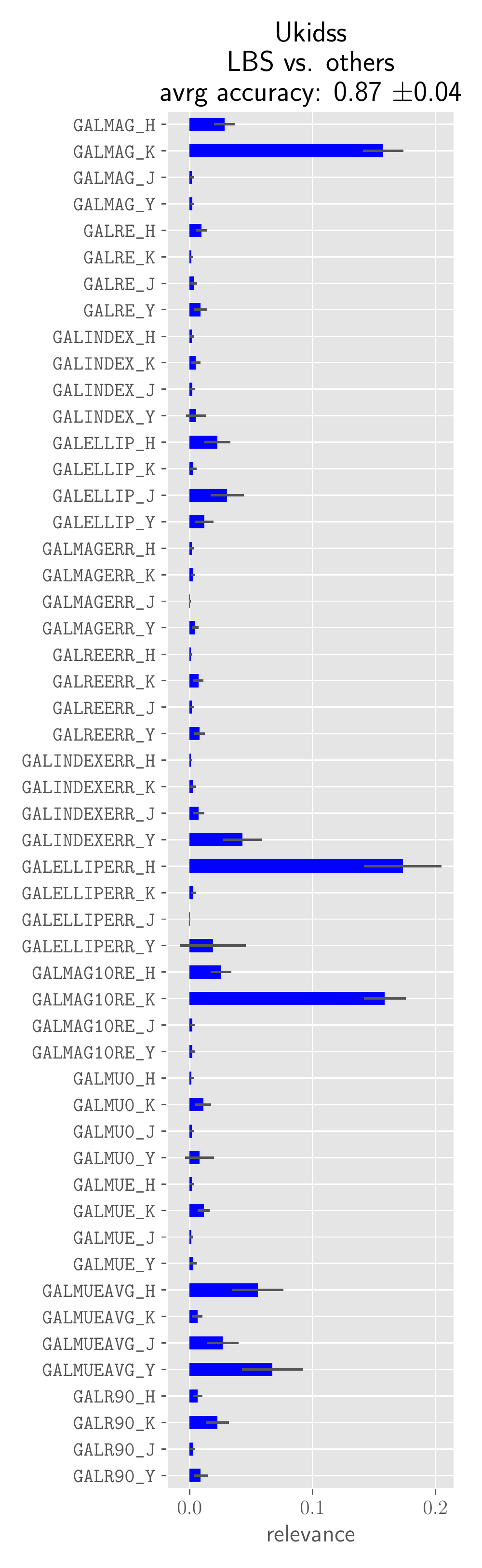}
        \label{fig:ukidss_LBS}
    \end{subfigure}
    \caption{
    Feature relevances as determined by GMLVQ for the SersicCatUKIDSS sample.
    \textcolor{black}{Same note applies here as to Fig.~\ref{fig:lambdar}.}
 }
    \label{fig:ukidss}
\end{figure}

\afterpage{%
        \thispagestyle{empty}
\begin{figure}
    \centering
        \includegraphics[width=1.1\textwidth]{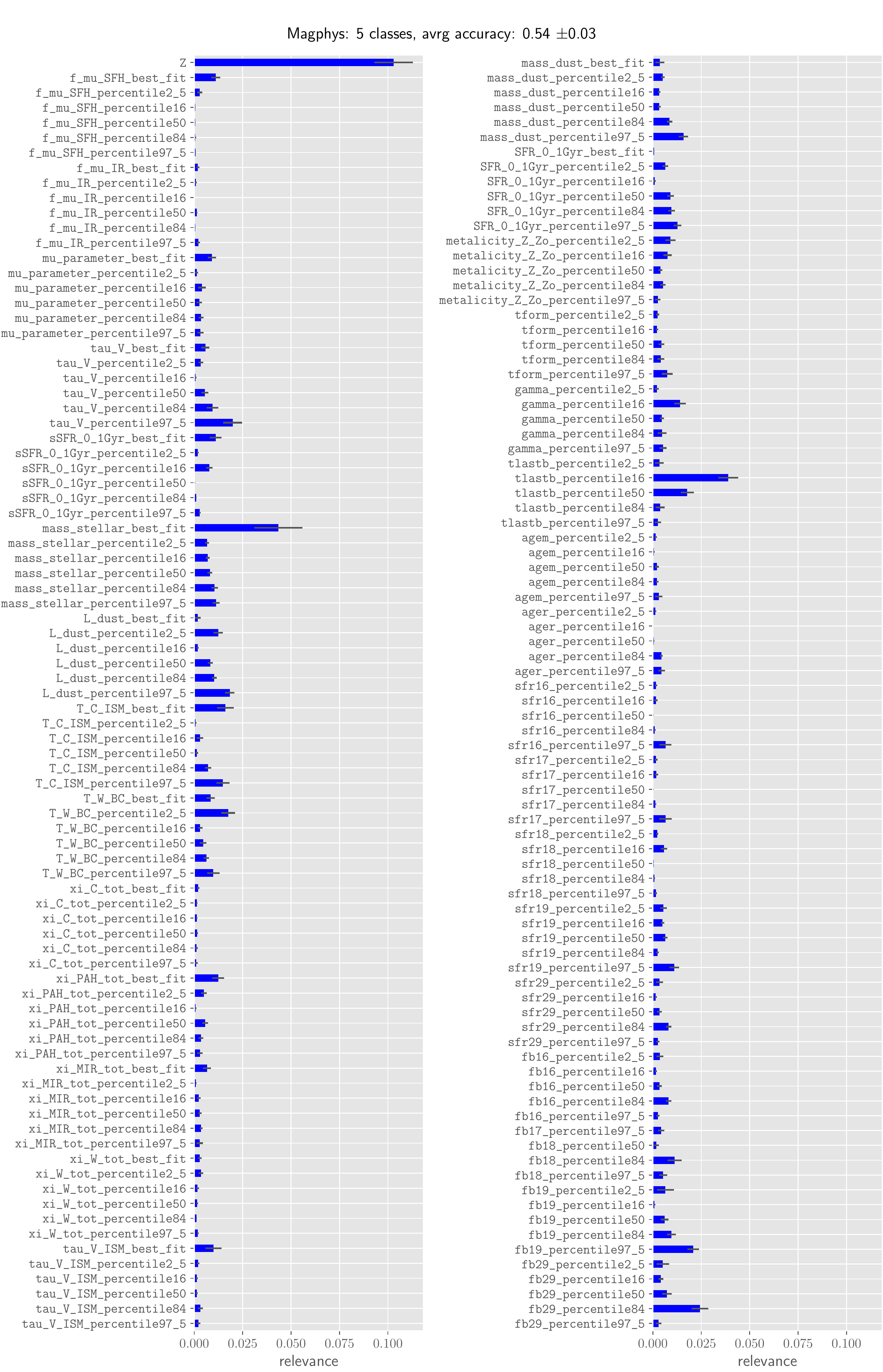}
    \caption{5 class problem: Feature relevances as determined by GMLVQ for the MagPhys sample.
     \textcolor{black}{Same note applies here as to Fig.~\ref{fig:lambdar}.}
 }
        \label{fig:magphys_5class}
    \end{figure}
    }
    
\afterpage{%
\thispagestyle{empty}
\begin{figure}
        \includegraphics[width=1.1\textwidth]{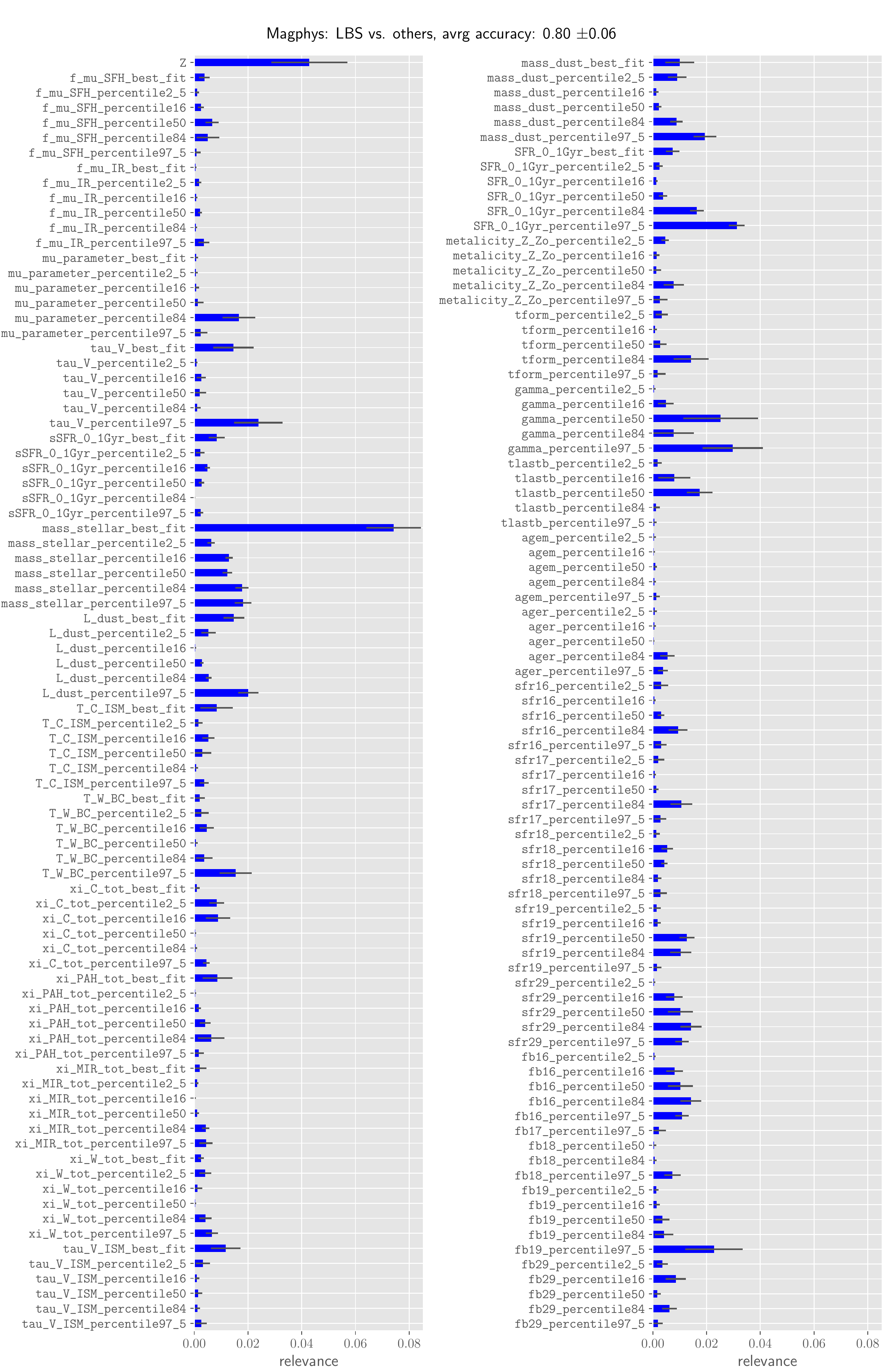}
    \caption{LBS vs. others: Feature relevances as determined by GMLVQ for the MagPhys sample. 
     \textcolor{black}{Same note applies here as to Fig.~\ref{fig:lambdar}.}
 }
    \label{fig:magphys_LBS}
\end{figure}
}
\FloatBarrier

\section*{References}

\bibliography{ref}

\end{document}